\documentclass[prd,twocolumn,showpacs,superscriptaddress]{revtex4}
\usepackage{epsfig}
\usepackage{graphicx}
\usepackage{amsmath}

\newcommand{\Tr}{{\rm \,Tr\,}}
\newcommand{\bi}{\bf}
\newcommand{\beq}{\begin{equation}} 
\newcommand{\eeq}{\end{equation}}
\newcommand{\beqa}{\begin{eqnarray}} 
\newcommand{\eeqa}{\end{eqnarray}}
\newcommand{\nhat}{\hat{\bf n}}
\newcommand{\threej}[6]{\left(\begin{array}{ccc}#1 & #2 & #3 \\ #4 & #5 & #6\end{array}\right)}

\newcommand{\comment}[1]{}

\begin{document}

\title{Correlation of CMB with large-scale structure: II. Weak lensing}

\author{Christopher M. Hirata}
\email{chirata@tapir.caltech.edu}
\affiliation{Caltech M/C 130-33, Pasadena, California 91125, USA}

\author{Shirley Ho}
\affiliation{Department of Astrophysical Sciences, Peyton Hall, Princeton University, Princeton, New Jersey 08544, USA}

\author{Nikhil Padmanabhan}
\affiliation{Lawrence Berkeley National Labs, 1 Cyclotron Road MS 50R-5032, Berkeley, California 94720, USA}

\author{Uro\v s Seljak}
\affiliation{Institute for Theoretical Physics, University of Zurich, 8057 Zurich, Switzerland}
\affiliation{International Center for Theoretical Physics, 34014 Trieste, Italy}
\affiliation{Department of Physics, University of California at Berkeley, Berkeley, California 94720, USA}

\author{Neta A. Bahcall}
\affiliation{Department of Astrophysical Sciences, Peyton Hall, Princeton University, Princeton, New Jersey 08544, USA}

\date{January 4, 2008}

\begin{abstract}
We investigate the correlation of gravitational lensing of the cosmic microwave background (CMB) with several tracers of large-scale structure, 
including luminous red galaxies (LRGs), quasars, and radio sources.  The lensing field is reconstructed based on the CMB maps from the Wilkinson 
Microwave Anisotropy Probe (WMAP) satellite; the LRGs and quasars are observed by the Sloan Digital Sky Survey (SDSS); and the radio sources are 
observed in the NRAO VLA Sky Survey (NVSS). 
Combining all three large-scale structure samples, we find evidence for a positive 
cross-correlation at the $2.5\sigma$ level ($1.8\sigma$ for the SDSS samples and $2.1\sigma$ for NVSS); the cross-correlation amplitude is $1.06\pm 
0.42$ times that expected for the WMAP cosmological parameters.  
Our analysis extends other recent analyses in that we carefully determine 
bias weighted redshift distribution of the sources, which is needed for a 
meaningful cosmological interpretation of the detected signal. 
We investigate contamination of the signal by Galactic emission, extragalactic radio and infrared sources, thermal and kinetic Sunyaev-Zel'dovich 
effects, and the Rees-Sciama effect, and find all of them to be negligible.
\end{abstract}


\maketitle

\section{Introduction}

Great progress has been made in cosmology in the past several years, in large part due to measurements of the anisotropies of the cosmic microwave 
background (CMB).  Most recently, the Wilkinson Microwave Anisotropy Probe (WMAP) experiment has generated high-resolution, all-sky maps of the CMB 
\cite{2007ApJS..170..288H,2007ApJS..170..335P}.  While a great deal of attention has been given to the ``primary'' anisotropies in the CMB imprinted at 
$z\sim 10^3$, experiments such as WMAP are also sensitive to secondary anisotropies caused by electron scattering or gravitational potentials at lower 
redshifts.  Among these secondary effects are the thermal and kinetic Sunyaev-Zel'dovich (tSZ/kSZ) effects, the integrated Sachs-Wolfe (ISW) effect, 
and gravitational lensing.  These secondary anisotropies are important for two reasons: first, they can act as ``foregrounds'' for primary CMB studies 
if they are not adequately modeled; and second, they provide information about the growth of structure at low redshift. Since the secondary 
anisotropies are subdominant on the large angular scales observed by WMAP, they are most easily detected by cross-correlation with large scale 
structure (LSS). Several groups have cross-correlated WMAP data with LSS tracers in order to study the ISW and tSZ secondaries and extragalactic point 
sources \cite{1998NewA....3..275B, 2003MNRAS.346..940D, 2003ApJ...597L..89F, 2003astro.ph..7335S, 2004Natur.427...45B, 2004ApJ...608...10N, 
2004MNRAS.347L..67M, 2004MNRAS.350L..37F, 2004PhRvD..69h3524A, 2005PhRvD..72d3525P, 2006MNRAS.365..171G, 2006MNRAS.365..891V, 2006MNRAS.372L..23C, 
2006PhRvD..74d3524P, 2006PhRvD..74f3520G, 2007MNRAS.376.1211M, 2007MNRAS.377.1085R}.

Lensing of the CMB is a secondary anisotropy that has attracted considerable attention because of its potential to probe the matter distribution at 
intermediate redshifts $z\sim O(1)$ \cite{1999PhRvD..59l3507Z, 2002ApJ...574..566H, 2003ApJ...590..664S, 2003PhRvL..91x1301K}.  Lensing of the CMB has 
several potential advantages as a cosmological probe relative to using lensing of galaxies: (1) it probes a redshift range inaccessible to galaxy 
lensing surveys since its ``source screen'' is at $z\approx 1100$; (2) its redshift is known to high accuracy; and (3) as a Gaussian random field the 
CMB does not suffer from intrinsic alignments.  Lensing of the CMB does however have its limitations: (1) the low signal-to-noise ratio with current 
data; (2) one cannot make tomographic measurements of the redshift dependence of the lensing signal without including external data sets 
\cite{2002PhRvD..65b3003H}; (3) foregrounds; and (4) instrumental systematics are a problem, especially for the lensing auto-power spectrum.

In the present paper, we aim to measure the weak gravitational lensing effect in the WMAP data.  We do this in cross-correlation with large-scale 
structure (LSS) in order to improve the signal-to-noise ratio.  We reconstruct the lensing deflection field \footnote{To be precise, we compute a 
filtered version of the lensing deflection field (\ref{eq:v}) to avoid technical problems due to sky cuts.  See Section~\ref{ss:cmblens} for details.} 
from WMAP using quadratic estimator methods \cite{2001ApJ...557L..79H}.  We then measure the cross-power spectrum of this deflection field with 
luminous red galaxy (LRG) and quasar maps obtained from the Sloan Digital Sky Survey (SDSS) and with the radio source maps from the NRAO VLA Sky Survey 
(NVSS). The detailed sample definitions, an analysis of the ISW effect, and parameter constraints are presented in a companion paper (``Paper I'' 
\cite{paperi}).

There have been two previous searches for the lensing effect on the CMB in cross-correlation.  Hirata {\em et~al.} \cite{2004PhRvD..70j3501H} used the 
the first-year WMAP data release and a smaller sample of LRGs from the SDSS.  They found a nondetection: $1.0\pm 1.1$ times the expected signal.  More 
recently, Smith {\em et~al.} \cite{2007PhRvD..76d3510S} used the third-year WMAP data and the NVSS radio sources (although with different selection 
cuts than those 
used here); they find a $3.4\sigma$ result ($1.15\pm 0.34$ times the expected signal, although their expected signal is 15\%\ larger due to a 
different bias, redshift distribution, and fiducial cosmology).  The analysis presented here draws heavily on that of Hirata 
{\em et~al.} \cite{2004PhRvD..70j3501H}, but benefits from: (i) reduced instrument noise due to the longer integration time in the 3-year WMAP data 
release; (ii) a larger sky area available in the SDSS; (iii) the inclusion of SDSS quasars and NVSS sources in addition to the LRGs; and (iv) improved 
treatment of tSZ and point source contamination. The current analysis also 
extends these previous analyses in that we put a lot of effort in accurate 
determination of the bias weighted redshift distribution, using 
cross-correlation information with other galaxy samples. 
Even though this is not required for claiming a detection, it 
is needed for any cosmological interpretation of the signal and 
thus for any consistency check of cosmology based on this analysis. 

The analysis of this paper will be mostly based on a ``fiducial'' cosmology, which is the 3-year WMAP-only best-fit 6-parameter cosmology 
\cite{2007ApJS..170..377S} from the point in the WMAP Markov chain with the highest likelihood.  The basic objective is to measure the strength of 
the lensing signal, determine its statistical error and sensitivity to systematics, and establish whether it is consistent with expectations for the 
fiducial cosmology.  We explore the effects of cosmological parameters on the lensing signal by running a Markov chain in Paper I.  The only 
place where we vary cosmological parameters is in some of the extragalactic foreground tests to show that even extreme variations in the 
assumed cosmology do not affect the basic conclusion that the foregrounds are negligible.  The parameter sets we will use are as follows.  The 
``fiducial'' (WMAP) cosmology is flat with $\Omega_bh^2=0.0222$, $\Omega_mh^2=0.1275$, $h=0.727$, $\sigma_8=0.743$, and $n_s=0.948$. 
The ``high-$\Omega_m$'' is an extreme model that maintains the angular diameter distance to the 
CMB and keeps $\Omega_bh^2$, $\Omega_mh^2$, $n_s$, and the primordial power spectrum the same but increases $\Omega_m$ until
there is no cosmological constant
($\Omega_\Lambda=0$, $\Omega_m=1.33$, $h=0.31$).  The ``high-$\sigma_8$'' model is the WMAP+SDSS vanilla model from Tegmark {\em et~al.} 
\cite{2004PhRvD..69j3501T} and has 
$\Omega_bh^2=0.0232$, $\Omega_mh^2=0.1454$, $h=0.695$, $\sigma_8=0.917$, and $n_s=0.977$.  The last model would give a large tSZ effect since it 
depends strongly on the normalization.  Of these models we consider the fiducial and high-$\sigma_8$ models to be viable, while the high-$\Omega_m$ 
model is already strongly ruled out by supernovae, large-scale structure, and measurements of the Hubble constant \cite{2001ApJ...553...47F,
2004ApJ...607..665R, 2004PhRvD..69j3501T, 2006A&A...447...31A} and is 
included only to make the point that the foreground analysis is insensitive to cosmology.

The outline of this paper is as follows: the theoretical predictions for lensing of the CMB are reviewed in Section~\ref{sec:theory}. The WMAP and 
SDSS data, including the LSS samples used, are described in Section~\ref{sec:data}.  The analysis methodology is presented in 
Section~\ref{sec:analysis}. We present our results in Section~\ref{sec:results} and systematic error estimates in Section~\ref{sec:sys}.  
Extragalactic foregrounds deserve special consideration and occupy Section~\ref{sec:exf}. We conclude in Section~\ref{sec:conc}.
The appendices cover the description of point source contamination in bispectrum language (Appendix~\ref{app:cont}); the halo model description of the 
bispectrum (Appendix~\ref{app:bs}); the kSZ and Rees-Sciama foregrounds (Appendix~\ref{app:efg}); cross-correlations of different foreground components 
(Appendix~\ref{app:corr}); and the weak lensing likelihood function (Appendix~\ref{app:lf}).

\section{Theory}
\label{sec:theory}

Gravitational lensing re-maps the primary CMB signal according to the 
relation
\beq
T(\nhat) = \tilde T(\nhat+{\bi d}(\nhat))
\label{eq:defl}
\eeq
Here $T(\nhat)$ is the observed CMB temperature in direction $\nhat$, 
$\tilde T$ is the unlensed (primary) CMB temperature, and ${\bi d}$ is the 
lensing deflection angle.  In the case where the lensing is adequately 
described by a single deflection (the ``Born approximation''), one may 
define a lensing potential $\Phi$ and convergence $\kappa$ satisfying 
${\bi d}=\nabla\Phi$ and $\kappa=-\nabla^2\Phi/2$, either one of which 
contains the full information in the lensing field 
\cite{2001PhR...340..291B}.  The convergence is given by
\beq
\kappa = 4\pi G\bar\rho_0\int_0^{\chi_{\rm CMB}}
(1+z)\sin^2_K\chi
(\cot_K\chi-\cot_K\chi_{\rm CMB})
d\chi,
\eeq
where $\bar\rho_0$ is the mean density of the universe
today, $\chi$ is the comoving radial distance, and $\sin_K\chi$ and $\cot_K\chi$ are
the sinelike and cotangentlike functions ($\chi$ and $1/\chi$ in a flat universe).
The correlation of the galaxy density with the convergence field can be 
determined by the Limber equation, which yields
\beqa
C_l^{g\kappa} \!\!&=&\!\! \frac{4\pi G\bar\rho_0}
{\int (dN/d\chi) d\chi}
\int \frac{dN}{d\chi} b_g (\cot_K\chi-\cot_K\chi_{\rm CMB})
\nonumber \\ && \!\!\times
(1+z)P(k=l/\chi) d\chi
\label{eq:clxk}
\eeqa
\cite{2004PhRvD..70j3501H}, where $dN/d\chi$ is the comoving distance distribution of the galaxies,
$P(k)$ is the matter power spectrum, and $b_g$ is the galaxy bias.
To measure the cross-spectrum $C_l^{g\kappa}$, we need to be able 
to reconstruct $\kappa$ from the WMAP data.  If we knew the primary CMB 
signal $\tilde T(\nhat)$, this would be a simple exercise. In the absence 
of this knowledge, we must rely on its statistical properties in order to 
do a lensing analysis.  The primary CMB is a Gaussian random 
field with power spectrum $\tilde C_l$, i.e.
\beq
\langle \tilde T_{l_1m_1}^\ast \tilde T_{l_2m_2} \rangle
= \tilde C_{l_1}\delta_{l_1l_2}\delta_{m_1m_2},
\eeq
where $\tilde T_{l_1m_1}$ is a multipole moment of the primary temperature 
field.  Expanding Eq.~(\ref{eq:defl}) to first order in ${\bi d}$, and 
working in harmonic space, one can show 
that the corresponding two-mode expectation value for the lensed CMB is
\beqa
\langle T_{l_1m_1}^\ast T_{l_2m_2} \rangle
\!\!&=&\!\! \tilde C_{l_1}\delta_{l_1l_2}\delta_{m_1m_2}
\nonumber \\ && \!\!
+ \sum_{LM} (-1)^{m_2}{\cal J}_{Ll_1l_2}
\nonumber \\ && \times
\threej{l_1}{l_2}{L}{-m_1}{m_2}{M}\kappa_{LM},
\label{eq:tt}
\eeqa
where the lensing coupling coefficient \cite{2004PhRvD..70j3501H} is
\beqa
{\cal J}_{Ll_1l_2} \!\! &=& \!\!
\frac{\sqrt{(2L+1)(2l_1+1)(2l_2+1)}}{L(L+1)\sqrt{4\pi}}
\threej{l_1}{l_2}{L}{0}{0}{0}
\nonumber \\ && \times
\bigl\{ [L(L+1)+l_1(l_1+1)-l_2(l_2+1)]\tilde C_{l_1}
\nonumber \\ && 
+ [L(L+1)-l_1(l_1+1)+l_2(l_2+1)]\tilde C_{l_2}
\bigr\}.\;\;
\eeqa
In Eq.~(\ref{eq:tt}), the first term is simply the unlensed expectation 
value, and the second term represents off-diagonal correlations (i.e. 
correlations between modes with different $l$ or $m$) induced by 
lensing.  The lensing reconstruction techniques that we will describe in 
Section~\ref{ss:cmblens} are based on an optimal weighting of these 
off-diagonal terms.

\section{Data}
\label{sec:data}

\subsection{Cosmic microwave background from WMAP}

We use the first three years of CMB temperature data \cite{2007ApJS..170..263J, 2007ApJS..170..288H, 2007ApJS..170..335P, 2007ApJS..170..377S}
from the WMAP satellite \cite{2003ApJ...583....1B, 2003ApJ...585..566P}
located at the Sun-Earth L2 Lagrange point.  WMAP carries a set of ten 
differencing assemblies (DAs) that measure the difference in microwave 
intensity between two points on the sky.  The satellite rotates through an 
interlocking scan pattern that allows each DA to build up a map of the 
entire microwave sky.  The DAs are designated K1, Ka1, Q1, Q2, V1, V2, W1, 
W2, W3, and W4, where the letters indicate the central frequency (K, Ka, 
Q, V, and W correspond to 23, 33, 41, 61, and 94 GHz respectively).  A new 
sky map is produced by each DA every year; since this analysis uses three 
years of WMAP data, this means there are 30 sky maps available.

The WMAP maps have been generated in the HEALPix\footnote{http://healpix.jpl.nasa.gov} pixelization system \cite{2005ApJ...622..759G} at resolution 9, 
which has $3\,145\,728$ pixels each covering a solid angle of 47.2 arcmin$^2$.  The actual resolution achieved in the maps is determined by the beam 
and hence varies with frequency, with the lower frequencies giving poorer resolution.  The lensing analysis relies on the high multipoles ($l>400$) 
that are only accessible to the V and W band DAs (which have 21 and 14 arcmin beams respectively), and hence we will almost exclusively use the maps 
from these bands.  The K, Ka, and Q band maps are never fed through the lensing pipeline because their large and highly elliptical beams do not 
preserve high-$l$ information.  The inclusion of these bands would only slightly reduce the total (CMB+noise) power spectrum $C_l+N_l$: for 3 bands 
(Q+V+W) the isotropically averaged $C_l+N_l$ is reduced by a maximum factor of 1.10 versus V+W only.  Including the K and Ka bands as well would 
improve this factor to 1.13 (relative to V+W).  This maximum is reached at $l\sim 400$ since at lower $l$ the sampling variance dominates and at higher 
$l$ the K, Ka, and Q band beam transfer functions are too small.  Since the lensing reconstruction is quadratic in temperature, the factor of 1.10 in 
temperature variance translates into at best a factor of 1.10 reduction in the standard deviation of $C_l^{g\kappa}$ (the improvement is even less than 
that if not all lensing information comes from $l\sim 400$ where the Q band gives the most improvement).  There is also the issue of point sources, 
which were already a major concern in our first-year analysis \cite{2004PhRvD..70j3501H} and would presumably become worse as the 3-year WMAP data 
reduces the noise at high $l$: because the lensing reconstruction is quadratic in temperature, flat-spectrum radio sources would produce 4.4 times as 
much contamination in Q band as in V and 19 times as much in Q as in W.  We therefore decided to use only V and W bands for the main lensing 
reconstruction, and use K, Ka, and Q only for foreground tests.

\subsection{Large scale structure from SDSS}

Two of the large-scale structure samples used in this analysis are photometric luminous red galaxies and quasars selected from SDSS imaging.  The SDSS 
drift-scans the sky in five bands ($ugriz$) \cite{1996AJ....111.1748F} under photometric conditions \cite{2000AJ....120.1579Y, 2001AJ....122.2129H} 
using a 2.5-meter optical telescope \cite{2006AJ....131.2332G}
with 3 degree field of view camera \cite{1998AJ....116.3040G}
located in New Mexico, USA \cite{2000AJ....120.1579Y}. The photometric and astrometric 
calibration of the SDSS and the quality assessment pipeline are described by Refs.~\cite{2002AJ....123.2121S, 2006AN....327..821T, 2007astro.ph..3454P, 
2003AJ....125.1559P,2004AN....325..583I}, respectively.  Bright galaxies \cite{2002AJ....124.1810S}, luminous red galaxies 
\cite{2001AJ....122.2267E}, and quasar candidates \cite{2002AJ....123.2945R} are selected from the SDSS imaging data for spectroscopic follow-up with a 
spectrograph connected to the same telescope \cite{2003AJ....125.2276B}.  We only use the imaging data since the spectroscopic galaxy sample covers 
only the very low redshifts (which are inefficient for CMB lensing) and the number density for the spectroscopic quasars is too low.  Additionally in 
angular cross-correlation there is no advantage to having precise redshifts.
The selection criteria for our LSS samples are related to, but distinct from, those for spectroscopic target 
selection. The SDSS has had seven major data releases \cite{2002AJ....123..485S, 2003AJ....126.2081A, 2004AJ....128..502A, 2005AJ....129.1755A, 
2004AJ....128.2577F, 2006ApJS..162...38A, 2007ApJS..172..634A}.

The use of SDSS LRGs and quasars as LSS tracers was driven by a compromise among several competing requirements.  First, there is a need for large sky 
coverage to improve statistics.  Second, one desires a sample with high number density and bias to minimize the degradation of the correlation 
coefficient due to Poisson noise.  Third, one would like the LSS sample to cover a broad range of redshifts, since the CMB is lensed by structures at 
all redshifts and use of an LSS sample with small $\Delta z$ implies a small correlation coefficient with the convergence field at $z=1100$.

The photometric LRGs observed by SDSS satisfy all three of our desirata; we have used the same selection criteria as were used in the previous lensing 
\cite{2004PhRvD..70j3501H} and angular clustering \cite{2007MNRAS.378..852P} analyses, although the sky coverage has increased by a factor of 2.  The 
photometric LRGs cover the redshift range $0.2<z<0.7$, which is significantly deeper than the SDSS spectroscopic galaxy and LRG samples.  However most 
of the CMB lensing arises from structures at higher redshift.  Therefore in this analysis we have also used the SDSS photometric quasars.  These can be 
seen from much larger distances than LRGs and cover the redshift range $z<2.7$, which broadens the overall redshift coverage of our LSS samples.
However their shot noise is greater: the LRGs have a comoving number density of $4\times 10^{-4}h^3\,$Mpc$^{-3}$, versus a maximum of
$10^{-5}h^3\,$Mpc$^{-3}$ for the quasars.
Of course, maximum theoretical signal-to-noise (though not necessarily the observed signal-to-noise) is achieved by combining the LRG and quasar 
constraints, and including the 
NVSS sources as described in the next section.

For this analysis we included the photometric SDSS data obtained between 1998 September 19 and 2005 June 11, and gridded in HEALPix resolution 10 due 
to the extensive small-scale structure in the survey boundaries.  The survey area after rejecting regions of high reddening or stellar density, bad 
seeing, and regions contaminated by bright stars is $2\,025\,731$ pixels (6641 deg$^2$) for the LRGs and $1\,842\,044$ pixels (6039 deg$^2$) for the 
quasars.  The difference is due to the more stringent reddening cut $E(B-V)<0.05$ used for the quasars, as compared to $E(B-V)<0.08$ for the LRGs.

The LRG and quasar samples in Paper I were sliced into thin redshift slices in order to measure the redshift evolution of the ISW effect.  The CMB 
lensing window function is broad and so we do not require high redshift resolution, thus for this paper both LRG samples in Paper I were grouped into a 
single LRG catalog, and both quasar samples were grouped into a single quasar catalog.  The number densities are 129 LRGs and 39 quasars per square 
degree. The details of the selection criteria and determination of sky cuts can be found in Paper I and will not be repeated here.

\subsection{NVSS sources}

The NVSS radio sources are useful for the lensing analysis because the survey covers a larger portion of the sky than SDSS ($27\,361$ deg$^2$) and is 
deeper than the LRGs: the median redshift is $z_{\rm med}\approx 1.0$.  The sample is identical to that considered in Paper I and its detailed 
construction will not be repeated here.  The main points are that the sources are selected to have flux $\ge 2.5\,$mJy in L-band (1.4$\,$GHz) and are 
unresolved at the Very Large Array (VLA) in the ``D'' configuration used for NVSS.  (The catalog goes down to 2.0$\,$mJy, however it is 50\%\ complete 
at 2.5$\,$mJy and there is a danger of spurious power if one goes to very low completeness levels without a detailed investigation of the noise 
properties of the survey.)  The survey covers the whole sky except for regions with $\delta<-40^\circ$.  We also removed heavily contaminated regions 
in the Galactic Plane ($|b|<10^\circ$) and regions contaminated by sidelobes from bright sources.  The survey details can be found in the technical 
paper by Condon et~al. \cite{1998AJ....115.1693C}.  These cuts accept $1\,104\,983$ sources.
As with the ISW analysis, the NVSS sources are treated as a single slice here because one does not 
have meaningful photometric redshifts.  Indeed, even the approximate determination of the redshift distribution was a significant challenge (see Paper 
I).

The bias-weighted redshift distribution $f(z)=b\,dN/dz$ for the NVSS sources is obtained by fitting a $\Gamma$-distribution to the cross-correlation of 
NVSS with the SDSS samples and with sources from the 2-Micron All Sky Survey (2MASS) \cite{2006AJ....131.1163S}.  The procedure is described in great 
detail in Paper I.  Note that we did not use the NVSS auto-power in our fits as it appears to contain spurious power whose angular spectrum is not 
precisely known.

\section{Analysis}
\label{sec:analysis}

\subsection{CMB lensing reconstruction}
\label{ss:cmblens}

The most straightforward method to correlate LSS with lensing of the CMB is to construct a map of the lensing convergence from the CMB data, and then 
compute the cross-power spectrum with the LSS data.  In practice the convergence map produced by existing reconstruction techniques is a highly 
nonlocal function of the CMB data, which represents a problem when cuts due to point sources or the Galactic plane are taken into account.  Therefore, 
we will pursue a slightly different strategy, namely to construct a vector field $\bi v$ from the CMB data that is correlated with the deflection 
field, but which only depends locally on the CMB temperature (i.e. there is very small dependence on points more than a few degrees away).  We
apply Galactic and point source masks to the field $\bi v$ instead of directly to the temperature map.
We then find the cross-power spectrum of $\bi v$ with each LSS sample by standard quadratic estimation techniques.  The same basic concept 
was used in \cite{2004PhRvD..70j3501H}, but some modifications have been made here so we describe the current version of the algorithm in detail.  The 
methodology is outlined in Section~\ref{sss:method}, and the choice of weight functions is determined in Section~\ref{sss:weights}.  The final 
construction 
of the lensing maps, including the masks, is described in Section~\ref{sss:maps}.  All real-space steps in the lensing reconstruction use HEALPix 
resolution 9.

\subsubsection{Methodology}
\label{sss:method}

We perform a reconstruction of the CMB lensing field using a quadratic estimator.  
Quadratic estimators have been proposed by many authors for lensing 
reconstruction from the CMB temperature \cite{1997A&A...324...15B, 
1999PhRvL..82.2636S, 2000PhRvD..62f3510Z, 2001ApJ...557L..79H}, CMB 
polarization \cite{2000PhRvD..62d3517G, 2002PhRvD..66f3008O, 
2003PhRvD..67h3002O, 2002ApJ...574..566H}, and diffuse high-redshift 21 cm 
radiation \cite{2004NewA....9..173C, 2004NewA....9..417P, 2006ApJ...653..922Z}.  For the 
reconstruction based on CMB temperature anisotropies such as those 
observed by WMAP, quadratic estimators provide similar signal-to-noise to 
more complicated likelihood-based methods \cite{2003PhRvD..67d3001H}, 
although this will not necessarily be the case for future CMB experiments 
that are sensitive to $B$-mode polarization \cite{2003PhRvD..68h3002H}.
The three key steps in constructing a quadratic estimator, as described in 
\cite{2001ApJ...557L..79H} are as follows.
\newcounter{steps}
\begin{list}{\arabic{steps}. }{\usecounter{steps}}
\item The temperature field 
is filtered to produce a weighted temperature map,
\beq
W_{lm} = \frac{T_{lm}({\rm observed})}{B_lC^{\rm wt}_l},
\label{eq:w}
\eeq
where $B_l$ is the beam transfer function (including both the physical 
beam and the Healpix pixel window function) and $C^{\rm wt}_l$ is a 
weighting function.  In the case considered by \cite{2001ApJ...557L..79H} 
where the noise is statistically isotropic, it is optimal to use $C^{\rm 
wt}_l = C_l+N_l$ where $N_l$ is the noise power spectrum; as discussed 
below statistically isotropic noise is not an appropriate assumption for 
WMAP.
\item The filtered temperature gradient is produced:
\beq
{\bi G}(\nhat) = \nabla \sum_{lm}\tilde C_lW_{lm}Y_{lm}(\nhat),
\label{eq:g}
\eeq
where $\tilde C_l$ is the unlensed CMB power spectrum (or one's best estimate of it).
\item A ``temperature-weighted gradient''
\beq
\tilde{\bi G}(\nhat)=W(\nhat){\bi G}(\nhat)
\label{eq:tg}
\eeq
is computed, and in the methodology of \cite{2001ApJ...557L..79H} this is 
filtered to produce a deflection angle or convergence map.
\end{list}
One can show that the end product of these manipulations is
\beqa
\tilde G_{lm} \!\!&=&\!\! (-1)^m
\sqrt{l(l+1)} \sum_{l'l''m'm''} \frac{{\cal J}_{ll'l''}}{4}
\nonumber \\ && \!\!\times
\threej{l}{l'}{l''}{-m}{m'}{m''} \frac{T_{l'm'}({\rm obs})T_{l''m''}({\rm obs})}{B_{l'}B_{l''}
C^{\rm wt}_{l'}C^{\rm wt}_{l''}}.
\label{eq:glm}
\eeqa
In the absence of lensing ($\kappa=0$), this will have expectation value zero because for $l\neq 0$ we either have $l'=l''$ and $m'=-m''$ (in which 
case either $l$ is even and the 3-$j$ symbol vanishes, or $l$ is odd and ${\cal J}_{ll'l''}=0$), or else $\langle T_{l'm'}({\rm obs})T_{l''m''}({\rm 
obs})\rangle=0$.  This is essentially a consequence of statistical isotropy.  A particular realization of the lensing field breaks 
statistical isotropy and induces off-diagonal terms in the covariance matrix
Eq.~(\ref{eq:tt}).  These add coherently to yield a contribution to $\langle\tilde G_{lm}\rangle$; at linear order in $\kappa$ this will be 
proportional to $\kappa_{lm}$ by rotational symmetry.

In practice, this methodology runs into several problems, and we make 
several changes in order to avoid them.  One problem is the leakage of the 
bright foregrounds in the Galactic plane into the survey region when 
Eqs.~(\ref{eq:w}) and (\ref{eq:g}) are implemented.  We solve this by setting 
the portion of the temperature map within the Galactic Kp2 cut to zero 
before computing Eq.~(\ref{eq:w}).  This results in artifacts near the Kp2 
boundary but as we will see most of the sky is uncontaminated.

The second 
problem is that for non-uniform noise such as that in WMAP, the above 
procedure turns out to be biased because the same instrument noise 
appears in $W(\nhat)$ and in ${\bi G}(\nhat)$, leading to a bias in 
$\tilde{\bi G}(\nhat)$.  [We can also see this from Eq.~(\ref{eq:glm}), since $T_{l'm'}$ and $T_{l''m''}$ contain noise correlations if the noise 
varies across the sky, which is certainly the case for WMAP.]  If the noise covariance matrix is known well this bias can be subtracted off.  The WMAP 
noise properties are clean enough that this is probably possible (especially if the resulting lensing map is to be used in cross-correlation).
However the simplest and most robust way to avoid this 
problem without being sensitive to possible errors in the noise covariance matrix is to use cross-correlations among the 18 V and W band maps 
obtained by WMAP.  Letting Greek indices $\alpha$, $\beta$, etc. denote 
the maps, we may replace Eq.~(\ref{eq:tg}) with
\beq
\tilde{\bi G}^{\alpha\beta}(\nhat)=\frac12[
W^\alpha(\nhat){\bi G}^\beta(\nhat) + W^\beta(\nhat){\bi 
G}^\alpha(\nhat)],
\label{eq:tg-use}
\eeq
where $W^\alpha(\nhat)$ represents the map produced by feeding the 
$\alpha$th WMAP map into the pipeline that constructs $W(\nhat)$.
This procedure provides $18\times17/2=153$ temperature-weighted gradient 
maps.
Since we have 18 maps there is little loss in using cross-correlations only (e.g. if the maps all had the same noise then the loss of 
signal-to-noise ratio ranges from 1 in the sampling variance limited case to $\sqrt{18/17}$ in the noise limited case); this is a 
small price to pay for eliminating the reliance on the detailed noise covariance.

The third problem with the standard quadratic estimator that comes up is 
that the filtering of $\tilde{\bi G}(\nhat)$ to obtain an unbiased 
estimator of $\kappa$ is highly nonlocal.  This is an issue because it 
spreads artifacts from point sources over the entire sky.  Our solution to 
this is to instead apply a Gaussian filter to $\tilde{\bi G}(\nhat)$
to obtain a filtered vector field ${\bi v}(\nhat)$:
\beq
v^{(\parallel,\perp)}_{lm} = e^{-l(l+1)\sigma_0^2/2}\tilde 
G^{(\parallel,\perp)}_{lm}.
\label{eq:v}
\eeq
Here $\parallel$ and $\perp$ represent the longitudinal (vector) and 
transverse (axial) multipoles, which are spin-1 analogues of the $E$ and 
$B$ multipoles for tensor fields.  The filtering scale $\sigma_0$ is set 
to $10^{-2}\,$radians (34 arcmin).  The fields ${\bi 
v}^{\alpha\beta}(\nhat)$ are suitable for cross-correlation studies since 
artifacts from point sources are local and can be masked.  Once they are 
constructed, the ${\bi v}^{\alpha\beta}(\nhat)$ are averaged together to 
make a final lensing map according to
\beq
{\bi v}(\nhat) = \frac{\sum_{\alpha\neq\beta} w_{\alpha\beta}
{\bi v}^{\alpha\beta}(\nhat)}
{\sum_{\alpha\neq\beta} w_{\alpha\beta}}.
\label{eq:vnhat}
\eeq
(The weights $w_{\alpha\beta}$ will be specified in Section~\ref{sss:weights}.)

\begin{figure}
\includegraphics[angle=-90,width=3.5in]{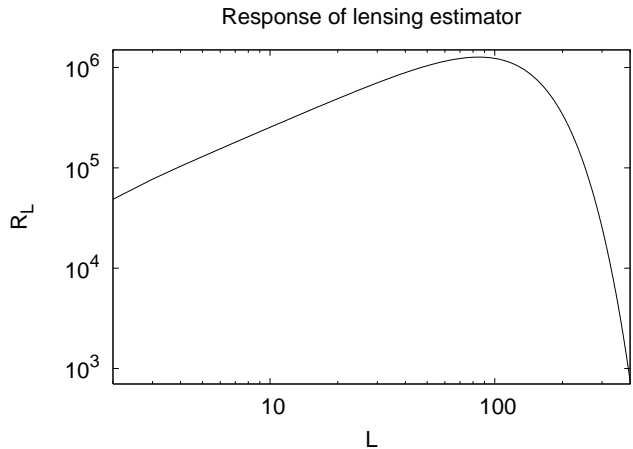}
\caption{\label{fig:resp}The response factor $R_l$ of Eq.~(\ref{eq:rl}).  This is the calibration relating the (mean of the) reconstructed map ${\bi 
v}$ to the underlying convergence factor $\kappa$.  Note that it depends on scale.}
\end{figure}

The use of the field ${\bi v}$ instead of a simple unbiased estimator for 
$\kappa$ means that some care must be taken in intepreting the LSS-${\bi 
v}$ cross-power spectrum.  Specifically, ${\bi v}$ is an unbiased 
estimator for some filtered version of $\kappa$ rather than $\kappa$ 
itself, and this filtering must be accounted for to obtain meaningful 
results.  Given a particular convergence field, the 
multipole moments of ${\bi v}^{\alpha\beta}$ can be shown to have 
expectation value
\beq
\langle v^{\alpha\beta(\parallel)}_{lm}\rangle = R_l\kappa_{lm},
\label{eq:vk}
\eeq
where the response factor $R_l$ is
\beq
R_l = \sum_{l'll'}
\frac{\sqrt{l(l+1)}}{4(2l+1)C_{l'}^{\rm wt}C_{l''}^{\rm 
wt}} {\cal J}_{ll'l''}^2e^{-l(l+1)\sigma_0^2/2}
\label{eq:rl}
\eeq
(cf. Eq.~17 of Ref.~\cite{2004PhRvD..70j3501H}).  This can be derived by plugging the expectation value from Eq.~(\ref{eq:tt}) into 
Eq.~(\ref{eq:glm}), using the 3-$j$ symbol orthogonality relations to collapse the sums over $m'$ and $m''$, and incorporating the
Gaussian factor from Eq.~(\ref{eq:v}).  This response factor is plotted in Fig.~\ref{fig:resp}.

\subsubsection{Weighting}
\label{sss:weights}

Finally, we note that the implementation of Eqs.~(\ref{eq:w}) and (\ref{eq:g}) 
requires an estimate of the unlensed CMB power spectrum $\tilde C_l$ and a 
choice of weighting function $C_l^{\rm wt}$.  For $\tilde C_l$ we use the 
{\sc CMBFast} \cite{1996ApJ...469..437S} prediction for the fiducial 
cosmology.  The ``actual'' $C_l$ estimates from WMAP could have 
been used instead, but this would have the undesirable feature of 
introducing a long-range dependence of $G(\nhat)$ on the CMB temperature, 
which is a problem if the Galactic plane has been removed.  We note that 
in any case an error in the assumed $\tilde C_l$ does not lead to any bias 
in the convergence field; rather it changes the response factor $R_l$ 
(\ref{eq:rl}).  This is equivalent to a (possibly scale-dependent) 
multiplicative or calibration bias in the lensing reconstruction, which
cannot give a spurious signal.

For $C_l^{\rm wt}$, we use an approximation to the total power spectrum 
(CMB+noise) obtained as follows.  Each of the 18 maps has a noise variance 
per pixel $\sigma_{\alpha,i}^2$ that depends on both the map and the pixel 
$i$.  In the limit where the noise is uniform and we use 
the Hu \cite{2001ApJ...557L..79H} version of the quadratic estimator, it 
would be optimal to define $C_l^{\rm wt}=C_l+N_l$.  We have thus chosen to 
write
\beq
C_l^{\rm wt} = 
\tilde C_l
+ (0.015655\;\mu{\rm K}^2)(B^{\rm V1}_l)^{-2}(1-l^2/1200^2)^{-4}
\label{eq:clwt}
\eeq
for $l\le 1200$ and $\infty $ for $l>1200$.
The second term in the $l\le 1200$ case is a good approximation to 
the effective noise in the $l\approx 400$ range where $N_l\approx C_l$.
Equation (\ref{eq:clwt}) has the advantage of having a cutoff in 
$C_l^{{\rm wt}\;-1}$ at high $l$, so Eq.~(\ref{eq:w}) can be computed by 
multiplication in harmonic space, and this cutoff is smooth, which avoids 
ringing when converted back to real space.  

Finally we come to the selection of the weights $w_{\alpha\beta}$.  If the 
WMAP beams were all the same and we were including the auto-correlations 
as well as cross-correlations in the lensing reconstruction -- i.e. if we 
included $\alpha=\beta$ terms in Eq.~(\ref{eq:vnhat}) -- then the optimal 
weight would be simply $w_{\alpha\beta}=1/(N_\alpha N_\beta)$, where 
$N_\alpha$ represents the noise variance in map $\alpha$.  For simplicity 
we choose to use this weighting scheme, where for $N_\alpha$ we have 
used the noise at the $l=400$ multipole (i.e. we take the 
mean-square noise per pixel and multiplied by the inverse-square beam 
$B_{400}^{-2}$).

\subsubsection{Masks and final lensing maps}
\label{sss:maps}

The lensing maps ${\bi v}$ have two major types of artifacts: one in the 
Galactic plane, induced by the Kp2 cut applied in Section~\ref{sss:method} 
and by foreground emission; and spurious features surrounding the point 
sources.  Both of these must be masked before proceeding.  The Galactic 
emission is treated by masking out all pixels within the Kp0 region, or 
within 5 degrees of its boundary.  This ``Kp05'' cut was also used in our 
first-year analysis \cite{2004PhRvD..70j3501H} and accepts $2\,064\,181$ 
HEALPix resolution 9 pixels ($f_{\rm sky}=0.656$).  We also remove any 
pixels centered within 2 degrees of a WMAP-detected source, as determined 
in the third-year catalog \cite{2007ApJS..170..288H}, which gives a final ``Kp052'' 
mask that accepts $1\,825\,036$ pixels ($f_{\rm sky}=0.580$).  Our main 
results will be shown with the Kp052 mask.

\begin{figure}
\includegraphics[width=3.4in]{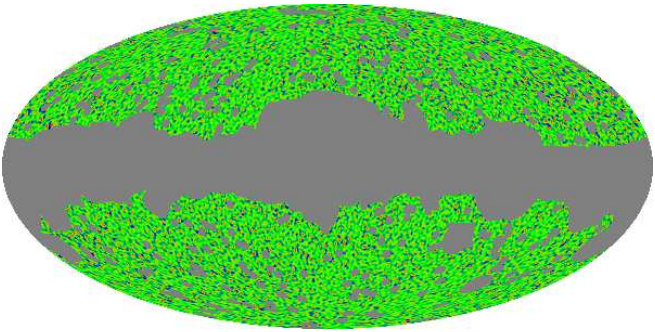}
\caption{\label{fig:cmblensmap}The CMB lensing map $\nabla\cdot{\bi v}^{(TT)}$ in Galactic Molleweide projection,
degraded to 40 arcmin resolution.  The gray region is rejected by the Kp052 mask.
The scale runs from $-2\times 10^{8}$ (black) to $+2\times 10^8$ (white).  The map is dominated by noise (both instrumental and due to the finite 
number of primary CMB modes available for the reconstruction).}
\end{figure}

In order to study the frequency dependence of the signal, we
construct individual-frequency maps ${\bi v}^{VV}$, ${\bi v}^{VW}$, and 
${\bi v}^{WW}$, which are obtained by applying Eq.~(\ref{eq:vnhat}) but 
restricting the averaging to products of maps in the specified bands.  For 
our weighting, we find that VV, VW, and WW contribute 26.35\%, 53.24\%, 
and 20.41\% respectively, i.e.
\beq
{\bi v}(\nhat) = 0.2635{\bi v}^{VV}(\nhat)
  + 0.5324{\bi v}^{VW}(\nhat) + 0.2041{\bi v}^{WW}(\nhat).
\label{eq:combo}
\eeq
The reconstructed lensing map is shown in Fig.~\ref{fig:cmblensmap}.

\subsection{Cross-correlation}

\subsubsection{Cross-spectrum estimation method}
\label{ss:cs}

We have cross-correlated the lensing maps ${\bi v}$ with the LRG and 
quasar maps $X$ using a quadratic cross-spectrum estimator.  
These methods have been extensively developed for both LSS and CMB
applications \cite{1997MNRAS.289..285H,1997MNRAS.289..295H, 
1997PhRvD..55.5895T}.  We write the LSS map as a vector of 
length $N_{\rm pix,LSS}$, and the CMB lensing map as a vector of length 
$2N_{\rm pix,CMB}$ (since it has two components).  The cross-spectrum is 
written in the form of a matrix,
\beq
C^{(\times)}_{ij} = \langle g_i v_j \rangle
= \sum_{lm} R_l C_l^{g\kappa} Y_{lm}^\ast(i) Y_{lm}^\parallel(j).
\eeq
Here $C^{(\times)}$ is an $N_{\rm pix,LSS}\times 2N_{\rm pix,CMB}$ matrix; 
$i$ is a pixel index in the LSS map; and $j$ encodes both a pixel index in 
the CMB lensing map and a unit vector on the celestial sphere ($\hat{\bi 
e}_\theta$ or $\hat{\bi e}_\phi$).  In the second equality we have written 
as shorthand $Y_{lm}(i)$ for the value of a scalar spherical harmonic at 
pixel $i$, and $Y_{lm}^\parallel(j)$ for the component of a vector 
spherical harmonic at the pixel and in the direction corresponding to 
index $j$.  This matrix is assumed to be a linear combination of 
templates, $C^{(\times)}_{ij} = \sum_A c^A P_{Aij}$, where the 
templates $P_{Aij}$ are known and their amplitudes $c^A$ are to be 
estimated.  We then construct the quadratic estimators
\beq
q_A = {\bi g}^T{\bf w}^{(g)}{\bf P}_A{\bf w}^{({\bi v})}{\bi v}
\eeq
and
\beq
F_{AB}= \Tr[
{\bf w}^{(g)}{\bf P}_A{\bf w}^{({\bi v})}{\bf P}_B^T],
\label{eq:qf}
\eeq
where ${\bf w}^{(g)}$ and ${\bf w}^{({\bi v})}$ are symmetric weight 
matrices ($N_{\rm pix,LSS}\times N_{\rm pix,LSS}$ and $2N_{\rm 
pix,CMB}\times 2N_{\rm pix,CMB}$ respectively).

The theory of quadratic estimation 
\cite{1997PhRvD..55.5895T} provides two key results concerning 
Eq.~(\ref{eq:qf}).  The first is that the 
quantities $\hat c^A=[{\bf F}^{-1}]^{AB}q_B$ are unbiased estimators of 
the $c^A$.  Note that this is true regardless of the (non-)Gaussianity or 
correlations between ${\bf g}$ and ${\bi v}$, and regardless of the 
choice of weight matrices (so long as ${\bf F}$ is invertible).  The 
second key result concerns optimality.  If
${\bi g}$ and ${\bi v}$ are Gaussian and weakly correlated, then 
the quadratic estimator is minimum-variance with the choice of weight 
matrices ${\bf w}^{(g)}=[{\bf C}^{(g)}]^{-1}$ and ${\bf w}^{({\bi v})}=[{\bf 
C}^{({\bi v})}]^{-1}$.  In our particular application the assumption of 
Gaussianity is violated, because ${\bi v}$ is constructed from products of 
the CMB temperature field and hence is itself non-Gaussian.  However, in 
the absence of better information, we use this optimality result to guide 
our choice of estimator.  The actual variance of the estimator will be 
given not by the matrix ${\bf F}$ (which in our context need not be the Fisher matrix) but rather by simulations (see 
Section~\ref{sec:stat}).

We have used two sets of templates ${\bf P}$ in the cross-correlation 
analysis.  The first set of templates uses band powers, i.e.
\beq
P_{Aij} = \sum_{l=l_{\rm min}(A)}^{l_{\rm max}(A)}
\sum_{m=-l}^l R_l Y_{lm}^\ast(i) Y_{lm}^\parallel(j),
\eeq
in which case the coefficients $c^A$ can be interpreted as the values of 
$C_l^{g\kappa}$ in the range $l_{\rm min}(A)\le l\le l_{\rm max}(A)$.  We 
use 13 bands covering the range from $2\le l\le 400$.  The other 
method is to use as a template the theoretical signal for the fiducial 
cosmology.

\subsubsection{Weights}

We require a weight matrix for the LRGs, the quasars, the NVSS sources, and the CMB lensing 
map ${\bi v}$.  The prescriptions below for the LRGs and quasars are 
believed to be a good approximation to the inverse-covariance matrix.  
For the CMB lensing map, the true covariance matrix is not known and ${\bf 
w}^{({\bi v})}$ should be thought of only as a heuristic weighting scheme.

For the LRGs, quasars, and NVSS sources, a preliminary power spectrum was measured in Paper I.  We have 
thus estimated the LRG covariance matrix as
\beq
C^{\rm(LRG)}_{ij} = \frac{\delta_{ij}}{\bar n}
  + \sum_{lm;\;l\le 400}
C_l^{\rm(LRG)} Y^\ast_{lm}(i)Y_{lm}(j),
\eeq
where the first term is the Poisson noise contribution, and
second term is the angular clustering.  The weight matrix is then ${\bf 
w}^{\rm(LRG)}=[{\bf C}^{({\rm LRG})}]^{-1}$.  The Poisson noise depends on 
the mean number of LRGs per pixel, $\bar n=0.424$.  Note that $\bar n$ is 
not allowed to depend on the pixel number, as this would place more weight 
on overdense regions and thus potentially bias the results.  The 
clustering contribution to the weight matrix is cut off at $l>400$, where 
Poisson noise dominates the LRG power spectrum, and we set 
$C_0=C_1=10^{-2}$ in order to reject the monopole and dipole.  A similar method was applied to the quasars and NVSS sources.

For the CMB lensing map, we followed a similar prescription for 
constructing a covariance matrix ${\bf C}^{({\bi v})}$ using the power 
spectrum of ${\bi v}$ taken from a simulation.  However there are three 
differences that appear when handling ${\bi v}$ instead of an LSS map.  
First, ${\bi v}$ is a vector so there are both longitudinal and transverse 
power spectra.  Secondly, the power spectrum of ${\bi v}$ is in fact 
variable over the sky, with higher values at low ecliptic latitude where 
the WMAP maps are more noisy.  Thirdly, ${\bi v}$ does not have ``shot 
noise'' on small scales, which makes the covariance matrix ${\bf C}^{({\bi 
v})}$ nearly singular.  We can work around all of these problems by 
writing
\beqa
C^{({\bi v})}_{ij,{\rm approx}} \!\!&=&\!\! \frac{\cal N}\Omega\delta_{ij}
 + \psi_i\psi_j\sum_{lm}
[C_l^{\parallel({\bi v})}-{\cal N}]
Y^{\parallel\ast}_{lm}(i)Y^\parallel_{lm}(j)
\nonumber\\&&\!\!
 + \psi_i\psi_j\sum_{lm}
[C_l^{\perp({\bi v})}-{\cal N}]
Y^{\perp\ast}_{lm}(i)Y^\perp_{lm}(j).
\label{eq:cva}
\eeqa
The use of both longitudinal ($\parallel$) and transverse ($\perp$) modes 
is necessary since ${\bi v}$ is a vector.  The factors $\psi_i$ 
attempt to weight different parts of the sky depending on their noise.  
Finally, the first term has been introduced to
ensure that ${\bf C}$ has no eigenvalues less than ${\cal 
N}/\Omega$, where the noise floor is ${\cal N}=2\times 10^5$ and the pixel 
area is $\Omega=4.0\times 10^{-6}$.  The sum over $l$ runs 
only over those values for which $C_l^{\parallel({\bi v})}>{\cal N}$; this 
is $l\le 265$ for the longitudinal modes and $l\le 267$ for the transverse 
modes.  (In these cases we take out the noise floor ${\cal N}$.)
The CMB lensing weight is then ${\bf w}^{({\bi v})}=[{\bf 
C}^{({\bi v})}_{\rm approx}]^{-1}$.

We also need to select the factors $\psi_i$, which control the relative 
weighting of different regions of sky.  Since ideally we would like 
Eq.~(\ref{eq:cva}) to be the true covariance matrix of ${\bi v}$, it is 
desirable to have $\psi_i^2$ proportional to the local power spectrum of 
${\bi v}$, which is highest near the Ecliptic.  We therefore divide the 
sky into several regions based on ecliptic latitude $\beta$, and computed 
pseudo-$C_l$ power spectra of $\bi v$ (again taken from a simulation 
rather than the real data in order to guard against possible correlations 
of the data with the weight matrix) in each.  It is found that these 
power spectra are $\sim 25\%$ higher than the average near the Ecliptic, 
and $\sim 25\%$ lower than average near the ecliptic poles.  Therefore we 
choose the heuristic weighting factors
\beq
\psi_i = \sqrt{1.25-0.50|\sin\beta_i|}.
\eeq
We have not chosen to do a more sophisticated weighting (e.g. more 
general dependence on $\beta$, or allowing the weighting factors to depend 
on $l$), noting that the quadratic estimator is unbiased regardless of the 
choice of ${\bf w}^{({\bi v})}$.

The ${\bf C}^{-1}$-type operations were performed using the 
conjugate-gradient algorithm, which was unpreconditioned for ${\bi v}$ and 
which used the preconditioner in Appendix B of \cite{2004PhRvD..70j3501H} 
with $l_{\rm split}=64$ for $g$.  The traces to obtain $F_{AB}$ are 
performed via the $Z_2$ stochastic trace method 
\cite{2003NewA....8..581P}.

\subsection{Statistical errors}
\label{sec:stat}

Simulations are frequently used in CMB studies in order to (i) estimate 
error bars; (ii) verify that the analysis pipeline correctly reconstructs 
simulation inputs such as maps or power spectra; and (iii) determine the 
effects of possible systematic errors in the data.  We use 
simulations for all three of these purposes, although only (i) is 
considered here (ii and iii are considered in Section~\ref{sec:sys}).  
There 
are two major types of simulations, one with a circular beam and one with 
a toy model of an elliptical beam.  The elliptical beam simulation is 
believed to be of higher fidelity, but we do a circular-beam analysis in 
order to understand whether beam ellipticity is an important effect for 
this analysis.

The circular-beam simulations described here are similar to those of 
Ref.~\cite{2004PhRvD..70j3501H}, except that they now produce 18 V and W band 
maps (one for each of the 3 years and 6 DAs) instead of the 8 Q, V, and W 
band maps produced in \cite{2004PhRvD..70j3501H}.  Additionally, the 
updated 3-year beam transfer functions $B_l^\alpha$ of \cite{2007ApJS..170..263J} 
are used in place of the 1-year $B_l^\alpha$ values 
\cite{2003ApJS..148...39P}.  The simulation assumes a Gaussian unlensed 
CMB $\tilde T$ and convergence $\kappa$ with power spectra generated by 
{\sc CMBFast} \cite{1996ApJ...469..437S} for the fiducial cosmology.
(The convergence field we generate is Gaussian but based on the nonlinear power spectrum.
However in our analysis we will restrict to the linear regime, in which case
Gaussianity should apply.)
We also generate elliptical-beam simulations.  The methodology for 
these simulations is described in Section~IVC of 
Ref.~\cite{2004PhRvD..70j3501H} 
and will not be repeated here; it consists essentially of taking the 
$m=\pm 2$ spherical harmonic components of the beam map for each DA and 
rotating them through a toy model of the WMAP scan strategy.  We have of 
course updated the beam spherical harmonic coefficients based on the 
3-year WMAP beam maps \cite{2007ApJS..170..263J}.

\section{Results and signal amplitude}
\label{sec:results}

\begin{figure*}
\includegraphics[width=6.5in]{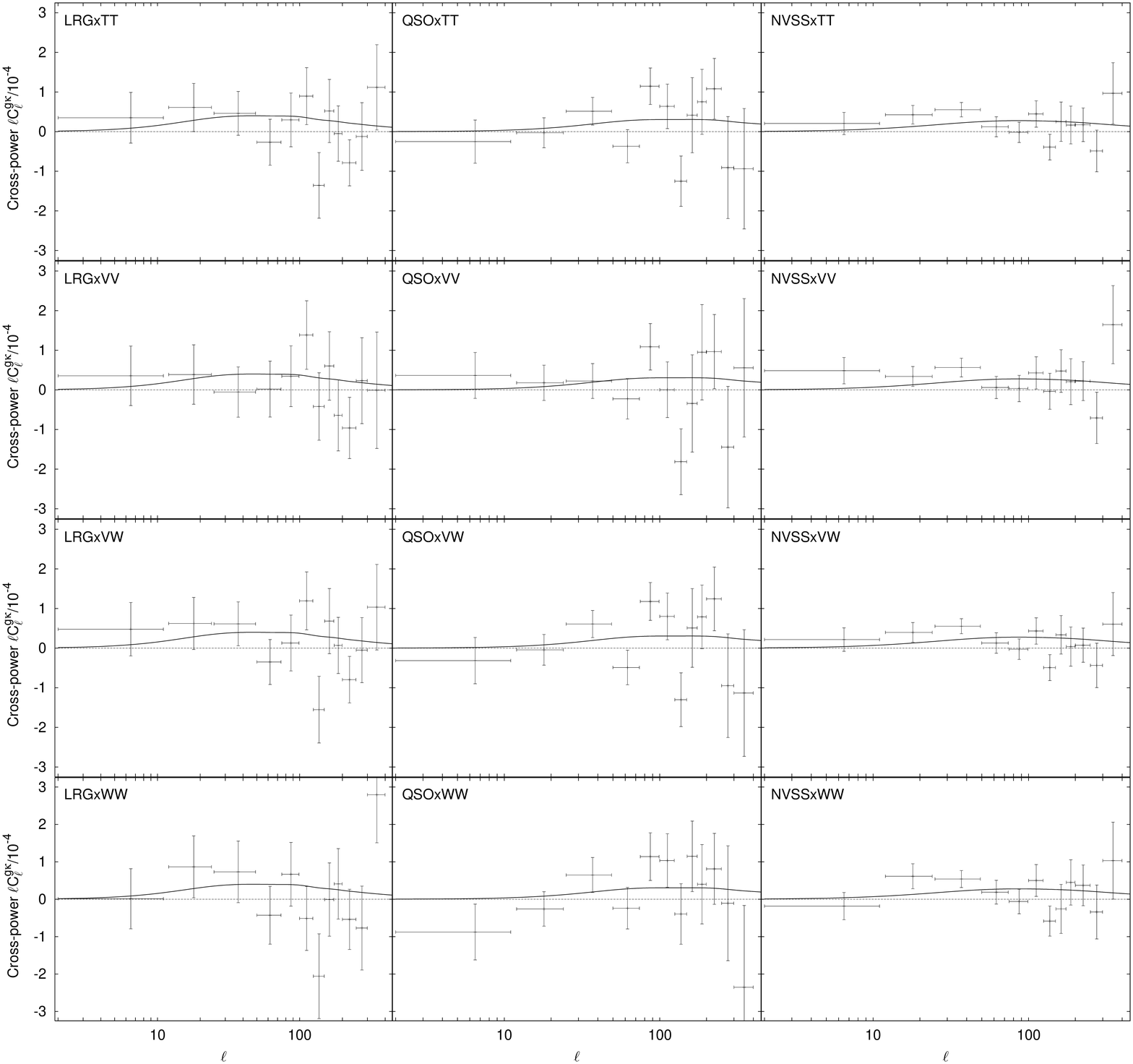}
\caption{\label{fig:signal}The LSS-convergence cross-spectra.  Each column represents a different large scale structure tracer (LRGs, quasars, and 
NVSS), and each row represents a different frequency combination (the first row, TT, is frequency-averaged; the other rows are VV, VW, and WW).  The 
horizontal dashed
line is zero, and the thick line shows the theoretical signal for the
fiducial cosmology and the best-fit LRG bias/redshift distribution (see Paper I).  The last data point in each plot is actually two $l$-bins that have 
been combined so that their error bars do not dominate the vertical scale of the plot.}
\end{figure*}

The cross-spectra between the large scale structure maps and the reconstructed convergence map are shown in Figure~\ref{fig:signal}.  The error bars 
are determined from cross-correlating the real LSS maps with 64 simulated elliptical-beam CMB lensing maps.

Of greatest interest to us is the signal amplitude and its statistical significance.  For each of the samples (LRGs, quasars, NVSS) we estimate the 
amplitude $A$ defined as the ratio of the observed signal to the theoretical signal in the WMAP best-fit cosmology.  That is, we fit
\beq
C_l^{g\kappa}({\rm obs}) = AC_l^{g\kappa}({\rm th})
\label{eq:adef}
\eeq
and determine the single parameter $A$.  Since for each sample there are several bins in $l$, we require a weight (inverse-covariance) matrix in order 
to do a least-squares fit and find the best $A$.  In principle the best covariance matrix for the $C_l^{g\kappa}$ is that obtained from the 
simulations.  However it is noisy and doing enough simulations to effectively eliminate this noise would be computationally prohibitive.  Therefore we 
have used the response matrix ${\bf F}^{-1}$ from the quadratic estimator procedure, Eq.~(\ref{eq:qf}).  Since the matrix ${\bf C}^{({\bi v})}$ used in 
this procedure is {\em not} the true covariance matrix, the matrix ${\bf F}^{-1}$ used here is not necessarily the true covariance.  For this reason it 
can only be used for weighting, and not for estimating the uncertainty in $A$.  For the latter we must use the simulations.  The amplitude is then 
determined from
\beq
A = \frac{\sum_{AB} [{\bf C}_w]^{-1}_{AB}C_{l_A}^{g\kappa}({\rm th})C_{l_B}^{g\kappa}({\rm obs})}
{\sum_{AB} [{\bf C}_w]^{-1}_{AB}C_{l_A}^{g\kappa}({\rm th})C_{l_B}^{g\kappa}({\rm th})}.
\label{eq:a}
\eeq
Here the sums are over the $l$-bins used in the fit, and $\{l_A,l_B\}$ are the values of $l$ at the centroids of the bin.  In cases where we do not 
use all of the $l$-bins, the weight matrices $[{\bf C}_w]$ are sub-blocks of ${\bf F}^{-1}$ corresponding to the $l$-bins used in the fit.  The 
theory used is linear theory, and we set $l_{\rm max}$ to correspond to $0.1 h\,$Mpc$^{-1}$ at the lowest quartile of the redshift distribution.  In 
equation form, $l_{\rm max} = k_{\rm max}/D_{A,25}$ where $k_{\rm max} = 0.1 h\,$Mpc$^{-1}$ and $D_{A,25}$ is the comoving angular diameter distance to 
the 25th percentile of the bias-weighted redshift distribution $f(z)$ (see Paper I).  This maximum value of $l$ is 107 (LRGs), 269 (quasars), and 186 
(NVSS).  Because we exclude entire $l$-bins if any range of multipoles is above the cutoff, this means we accept 5 bins ($l_{\rm max}=100$) for the 
LRGs, 10 bins ($l_{\rm max}=250$) for the quasars, and 8 bins ($l_{\rm max}=175$) for NVSS.
If we combine the LRGs, quasars, and NVSS, the weighting procedure of Eq.~(\ref{eq:a}) gives relative weights of 0.239, 0.395, and 0.365 for 
the three samples, respectively.

The results from this procedure are shown in Table~\ref{tab:results}; the error bars are again determined by cross-correlating the real LSS maps with 
the 64 simulated CMB lensing maps.  We have also shown in the table the results that are obtained by changing $k_{\rm max}$ from its fiducial value of 
0.1 to 0.05 or 0.15 $h\,$Mpc$^{-1}$.  As one can see these changes in the analysis result in $<1\sigma$ changes in the amplitude $A$.

The primary result of this paper is the standard procedure using the TT bands (i.e. the signal averaged over the V and W bands).  This is $A=+1.06\pm 
0.42$, i.e. formally a $2.5\sigma$ signal.  Note that the relative weighting of LRGs, QSOs, and NVSS is determined from the response matrix, 
Eq.~(\ref{eq:a}).  If we used weights from the simulation covariance instead the amplitude would instead be $+1.08\pm 0.42$, a negligible change.

\begin{table*}
\caption{\label{tab:results}The amplitude $A$ of the lensing cross-correlation signal normalized to that predicted for the WMAP cosmology 
(Eq.~\ref{eq:adef}).  The first four rows show the ``standard'' fit as described in the text, following rows show the consequences of modifying 
this procedure.}
\begin{tabular}{lcccccccccc}
\hline\hline
Fit type & & WMAP bands & & $A$ (LRGs) & & $A$ (QSOs) & & $A$ (NVSS) & & $A$ (combined) \\
\hline
Standard & & TT & & $+0.72\pm 0.76$ & & $+1.20\pm 0.73$ & & $+1.11\pm 0.52$ & & $+1.06\pm 0.42$ \\
Standard & & VV & & $+0.42\pm 0.85$ & & $+0.48\pm 0.83$ & & $+1.28\pm 0.64$ & & $+0.92\pm 0.48$ \\
Standard & & VW & & $+0.71\pm 0.78$ & & $+1.33\pm 0.76$ & & $+1.08\pm 0.54$ & & $+1.07\pm 0.44$ \\
Standard & & WW & & $+1.16\pm 1.19$ & & $+1.80\pm 0.91$ & & $+0.98\pm 0.66$ & & $+1.21\pm 0.54$ \\
\hline
$k_{\rm max}=0.05h\,$Mpc$^{-1}$ & & TT & & $+1.51\pm 1.12$ & & $+1.56\pm 0.80$ & & $+1.36\pm 0.60$ & & $+1.43\pm 0.50$ \\
$k_{\rm max}=0.15h\,$Mpc$^{-1}$ & & TT & & $+0.48\pm 0.57$ & & $+1.03\pm 0.73$ & & $+0.99\pm 0.52$ & & $+0.88\pm 0.39$ \\
\hline\hline
\end{tabular}
\end{table*}

We have also computed the mean values of $A$ obtained from the simulated CMB lensing maps; these are $+0.01\pm0.10$, $+0.08\pm0.09$, and 
$+0.03\pm0.07$.  
These are all consistent with zero at the $1\sigma$ level.  (We have quoted error bars on the mean of 64 simulations, which are a factor of 
$\sqrt{64}=8$ smaller than the error bars on the data.)
Since the LSS maps in our analysis are real instead of random, these mean values should be 
considered a test of whether the galaxy-convergence cross correlation that we observe is due to some feature of the galaxy map being aligned by chance 
with the CMB mask.

A signal that is formally $2.5\sigma$ may not be statistically significant if the noise distribution (value of $A$ under the null hypothesis of no 
lensing) is non-Gaussian.  We have therefore computed the skewness and kurtosis of the Monte Carlo values of $A$.  For the combined signal we find 
$\langle A^3\rangle/\langle A^2\rangle^{3/2} = +0.15\pm 0.48$ and $\langle A^4\rangle/\langle A^2\rangle^2 = 2.46\pm 0.61$, which for a Gaussian 
distribution should be 0 and 3 respectively.  Thus with the 64 simulations there is no evidence for any departure from Gaussianity.

\section{Systematic errors}
\label{sec:sys}

This section considers possible systematic errors resulting from instrumental or algorithmic effects, Galactic emission, and some general tests on the 
reconstructed map.  The extragalactic foregrounds require separate consideration, and discussion of them is deferred to Section~\ref{sec:exf}.

\subsection{Calibration and beam ellipticity}
\label{ss:calib}

We have tested the calibration of the lensing estimator using the simulations from Section~\ref{sec:stat}.  Aside from being an important test of the 
code, we note that the calibration of the lensing estimator could differ from unity for several reasons.  First, Eq.~(\ref{eq:vk}) was derived assuming 
the small-deflection limit, i.e. it is constructed only to first order in the deflection angle ${\bi d}$.  The simulation does not use any Taylor 
expansion in ${\bi d}$.  Secondly, there are a number of places in the reconstruction and cross-power estimation codes where interpolations, 
$l$-cutoffs, and iterative ${\bf C}^{-1}$-type operations are used, and it is important to test whether these have introduced calibration biases.  
Finally, the beam ellipticity causes the effective transfer function $B_l$ to depend not just on the magnitude of the wavenumber $l$ but also on its 
direction, which is not taken into account in constructing Eq.~(\ref{eq:vk}).

We have used simple simulations to estimate these effects.  In each case a Gaussian $\kappa$ map and corresponding lensed temperature maps were 
constructed, as described in Section~IVC of Ref.~\cite{2004PhRvD..70j3501H}.  These maps were beam-smoothed, and we added Gaussian uncorrelated noise 
(based on the WMAP $N_{\rm obs}$ values), which is accurate for the high multipoles where noise is significant.  They were then fed through the 
reconstruction pipeline to determine ${\bi v}^{TT}$.  A galaxy field was generated with the correct cross-spectrum $C^{g\kappa}_l$ by setting 
$g_{lm}=(C^{g\kappa}_l/C^{\kappa\kappa}_l)\kappa_{lm}$.  (In principle there is also an uncorrelated contribution to the galaxy field, i.e. we should 
be adding an additional contribution with power spectrum $C_l^{gg}-C_l^{g\kappa\,2}/C_l^{\kappa\kappa}$; however when computing cross-correlations this 
term will average to zero, and including it merely adds noise to the simulation results.  We used the theoretical $C_l^{g\kappa}/C_l^{\kappa\kappa}$ 
for these simulations, based on the bias and redshift distributions for the fiducial cosmology from Paper I.)  The input power spectra are those from 
the fiducial cosmology, and the output spectra $\hat C_l^{g\kappa}$ are fit to construct an amplitude $A$, just as with the real data. Since the input 
lensing amplitude is $A_{\rm in}=1$, the output $A_{\rm out}$ can be used as a test of the calibration of the lensing pipeline.  Ideally $A_{\rm 
out}/A_{\rm in}=1$.  The results from this test are given in Table~\ref{tab:calib} for both circular-beam and elliptical-beam simulations; the 
elliptical-beam case is closer to reality, but the circular-beam case is useful for separating out the possible contributions to the calibration error.  
Note that the calibration bias $A_{\rm out}/A_{\rm in}-1$ is within $2\sigma$ of zero in most cases, the exception being the case of the circular beam 
in cross-correlation with NVSS.  The implied calibration error for this case is $\sim 6$\%, which is still much smaller than the statistical error.

\begin{table}
\caption{\label{tab:calib}The calibration of the lensing estimator, i.e. the mean value of $A_{\rm out}/A_{\rm in}$ obtained from a set of 64 
simulations with the fiducial cosmology, and the $1\sigma$ error on this mean value.  For perfect calibration this factor is 1.}
\begin{tabular}{ccccccc}
\hline\hline
Beam & & \multicolumn{5}{c}{\mbox{$A_{\rm out}/A_{\rm in}$}} \\
type & & LRG & & QSO & & NVSS \\
\hline
Circular & & $0.930\pm 0.035$ & & $0.979\pm 0.026$ & & $1.059\pm 0.019$ \\
Elliptical & & $0.953\pm 0.038$ & & $0.959\pm 0.024$ & & $1.040\pm 0.021$ \\
\hline\hline
\end{tabular}
\end{table}

\subsection{Power spectrum of reconstructed map}
\label{ss:ps}

One test of the reconstructed map is to compare its power spectrum to 
simulations.  This test is useful because it is sensitive to any source of 
spurious power in the lensing reconstruction such as foregrounds, 
unmodeled beam ellipticity, or $1/f$ noise.  It is also important because 
the error bar on the LSS-convergence cross correlation depends on the 
convergence power spectrum (including noise and systematics), so it is 
essential to compare the measured convergence power spectrum to 
simulations in order to establish the reliability of the simulation-based 
error bars.

\begin{figure}
\includegraphics[width=7.2in,angle=-90]{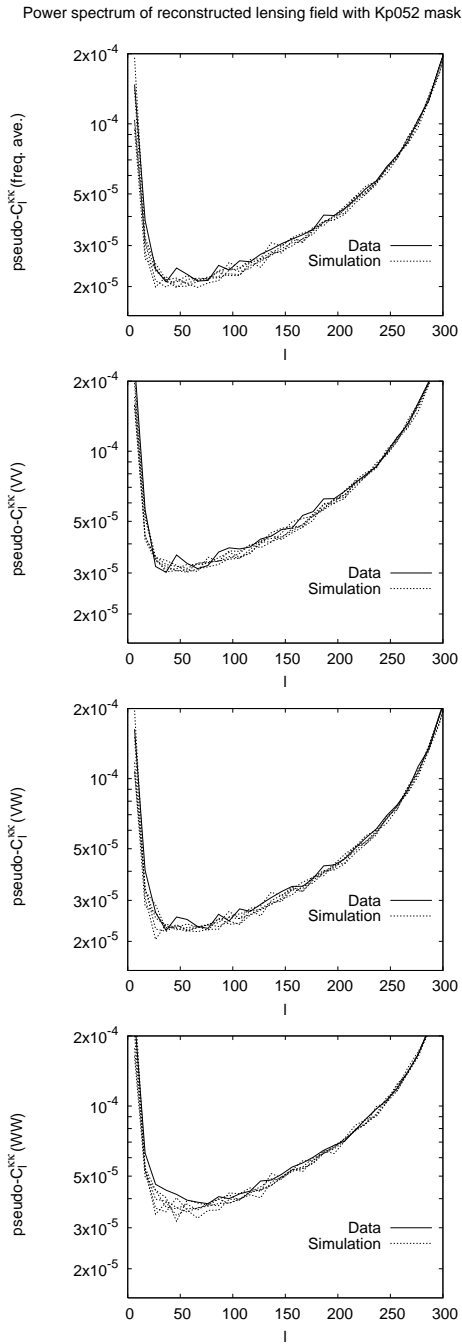}
\caption{The pseudo-$C_l$ power spectrum of the reconstructed lensing map 
(Eq.~\ref{eq:pcl}) for the Kp052 cut.  These power spectra are dominated by noise at all scales.
The top panel shows the 
power spectrum of the full reconstruction, whereas the other panels show 
the power spectra of the ${\bi v}^{VV}$, ${\bi v}^{VW}$, and ${\bi 
v}^{WW}$ lensing maps.  The solid line shows the data, and the dotted 
lines show five simulations.  Note that with the Kp052 sky cut the simulations 
reproduce the power spectrum of the data to within several 
percent.  Similar plots for the Kp05 cut show significant excess power.
\label{fig:psvec_mask0}}
\end{figure}

In Figure~\ref{fig:psvec_mask0}, we show the pseudo-$C_l$ power spectrum 
of 
the reconstructed lensing field with the Kp052 mask.  This power spectrum 
was derived via the 
formula
\beq
{\rm pseudo-}C^{\kappa\kappa}_l = 
\frac{\sum_{m=-l}^l\left| \int {\bi 
v}(\nhat)\cdot
{\bi Y}^{\parallel\ast}_{lm}(\nhat)\,d^2\nhat \right|^2}
{(2l+1)R_l^2f_{\rm sky}},
\label{eq:pcl}
\eeq
where ${\bi Y}^{\parallel\ast}_{lm}$ are the vector spherical harmonics.  
As can be seen from Eq.~(\ref{eq:vk}), this is simply the power spectrum 
of the longitudinal component of ${\bi v}$, divided by $R_l^2$ to convert 
it to a convergence power spectrum.  Note that since our objective here is 
simply to compare the amount of power in the data with the simulations, we 
have not done any noise subtraction, and we have divided by
$f_{\rm sky}$ instead of doing a full deconvolution of the Kp052 mask.  
The pseudo-$C_l$s in the figure have been binned in spacings $\Delta 
l=10$.  We have also shown this power spectrum for the 
individual-frequency maps ${\bi v}^{VV}$, ${\bi v}^{VW}$, and ${\bi 
v}^{WW}$; since the power spectrum of the convergence is noise-dominated, 
it is of course lowest in the frequency-averaged map ${\bi v}$.

For comparison, we repeated the same analysis for the Kp05 cut, i.e. 
without masking the point sources.  This resulted in excess power that is not present in the simulations; this is
especially the case in the ${\bi v}^{VV}$ map, which has the highest point
source contamination.  For this reason the Kp05 mask was not used in the rest of our analysis.

\subsection{90$^\circ$ rotation test}

\begin{figure*}
\includegraphics[width=6.5in]{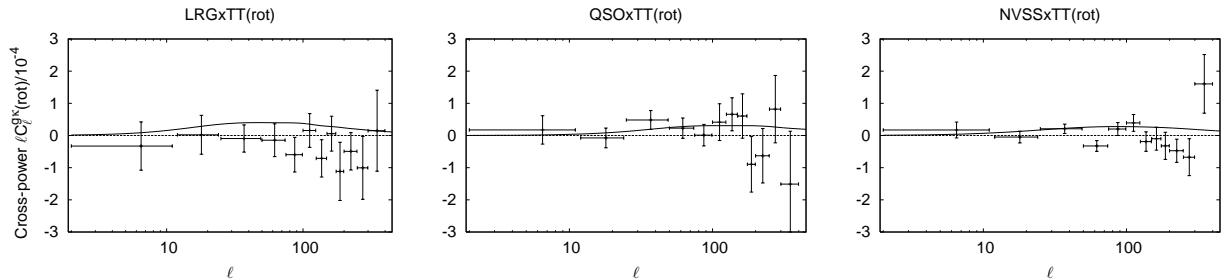}
\caption{\label{fig:cross-lrg90}The LSS-convergence cross spectrum $C_l^{g\kappa}$ obtained by rotating ${\bi v}$ 90 degrees before computing the 
cross-correlation.  This should be zero in the absence of systematics.  The three panels show the LRGs, quasars, and the NVSS sample.  The solid lines 
show the signal expected for the {\em unrotated} maps in the WMAP cosmology.  The last data point in each plot is actually two $l$-bins that have
been combined so that their error bars do not dominate the vertical scale of the plot.}
\end{figure*}

An important systematics test in lensing analyses using galaxies as sources is to rotate each galaxy ellipticity by 45 degrees and look for a lensing 
signal.  This is commonly known as the ``$B$-mode test'' since the rotation by 45 degrees interconverts the $E$-mode shear pattern produced by lensing 
and the $B$-mode pattern that should be zero in the absence of systematics.  For our analysis, the reconstructed CMB lensing field is a vector with 
spin 1, hence the analogous test here is to rotate ${\bi v}$ by 90 degrees and re-measure $C_l^{g\bi v}$.  This is done in Figure~\ref{fig:cross-lrg90} 
for each of our samples.  The $\chi^2$ for a null signal using the covariance matrix from rotated simulations is 7.85 for the LRGs, 14.55 for the 
quasars, and 17.32 for NVSS for 13 degrees of freedom. One can also fit for the amplitude $A_{90}$ of these rotated cross-power spectra using the 
theoretical lensing signal as a template and the same range of $l$ as for the main fit [analogous to Eq.~(\ref{eq:a}), except that this time one 
expects to get zero].  This gives $A_{90}=-0.61\pm0.72$ for the LRGs; $A_{90}=+0.77\pm0.62$ for the quasars; and $A_{90}=+0.16\pm0.33$ for NVSS, with 
the error bars determined from the rotated simulations.  These are indeed consistent with zero (the largest value is for the quasars which are 
$1.24\sigma$ from zero).

\subsection{Frequency dependence of signal}
\label{ss:frequency}

One simple but important test for the CMB lensing origin of the signal we observe is the frequency dependence.  If CMB lensing is the correct 
explanation then the signals in the VV, VW, and WW lensing maps will be identical (aside from instrument noise).  If instead the signal is due to some 
other effect, such as point sources correlated with the LSS tracer, then a different frequency dependence is expected.  In general for each spectrum we 
can construct the ratio of thermodynamic temperatures $R_{V/W}\equiv T(V)/T(W)$.  The lensing amplitudes $A$ obtained for each of the lensing maps 
should then be in the ratio $R_{V/W}:1:R_{V/W}^{-1}$ for the VV:VW:WW maps, respectively.  The most worrying foregrounds are flat-spectrum radio 
sources ($R_{V/W}=2.1$ for $\alpha=0$), infrared sources ($R_{V/W}=0.46$ for $\alpha=3.5$), and the thermal Sunyaev-Zel'dovich (tSZ) effect 
($R_{V/W}=1.16$).  By comparison, 
frequency-independent CMB fluctuations should have $R_{V/W}=1$.

\begin{figure}
\includegraphics[height=3.2in,angle=-90]{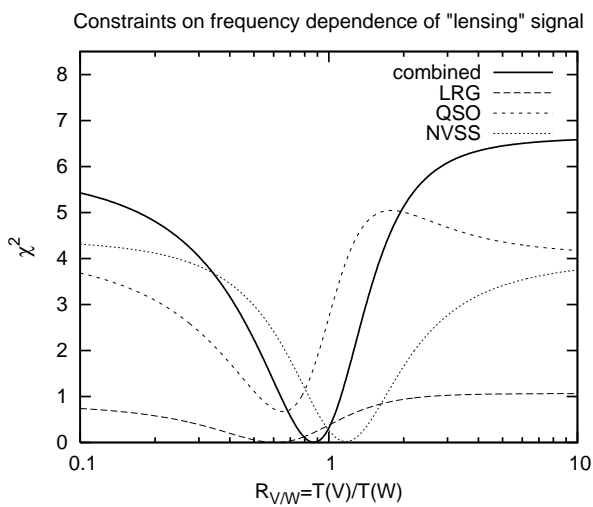}
\caption{\label{fig:fdplot}Constraints on the frequency dependence of the lensing signal.  The horizontal axis is the ratio of thermodynamic 
temperatures in the V and W bands, which is 1 for the CMB.  The vertical axis is the $\chi^2$ obtained by fitting the values of the lensing amplitude 
$A$ in the VV, VW, and WW frequency combinations.}
\end{figure}

The simplest way to set a constraint on $R_{V/W}$ is as follows.  For each LSS sample, we obtained in Section~\ref{sec:results} the amplitudes 
$A_{VV}$, 
$A_{VW}$, and $A_{WW}$, which form a 3-component vector ${\bf A}$, and their $3\times 3$ covariance matrix ${\rm Cov}_{\bf A}$ is known from the Monte 
Carlo simulations.  We also know that given $R_{V/W}$ the vector ${\bf A}$ should be parallel to the template vector ${\bf 
t}(R_{V/W})=(R_{V/W},1,R_{V/W}^{-1})$.  Then we construct the $\chi^2$:
\beq
\chi^2 = \min_y [{\bf A} - y{\bf t}(R_{V/W})]^T[{\rm Cov}_{\bf A}]^{-1}[{\bf A} - y{\bf t}(R_{V/W})].
\eeq
The $\chi^2$ curves for each sample (LRG, quasar, and NVSS) and the combined case are shown in Figure~\ref{fig:fdplot}.
For the combined case, we find $R_{V/W} = 0.87^{+0.74}_{-0.57}$ (2$\sigma$, determined by setting $\chi^2-\chi^2_{\rm min}=4$).

For each of the LSS samples used in this paper, the frequency dependence of the LSS-convergence correlation is consistent with a blackbody spectrum 
($R_{V/W}=1$).  In particular for the combined LSS sample, the frequency dependence disfavors the possibility that the signal is due to flat-spectrum 
radio sources at $2.3\sigma$.  With the frequency dependence tests alone, we cannot rule out the possibility that the signal is due to infrared 
sources or tSZ haloes; our constraints on these contaminants are discussed in Section~\ref{sec:exf}.

%\begin{table}
%\caption{\label{tab:fd}The values of the V:W band temperature ratios $R_{V/W}$ and their $1\sigma$ and $2\sigma$ confidence limits, as determined by 
%analyzing the 
%cross-correlation of large scale structure with the VV, VW, and WW lensing maps.  If foregrounds are negligible this ratio should be 1; if the entire 
%signal we observe is due to foregrounds, then $R_{V/W}$ should be the V:W band temperature ratio of the foreground.  In some cases no lower or upper 
%limit can be obtained due to the weakness of the signal; in these cases the limits are quoted as 0 or $\infty$.  The minimum $\chi^2$ has one degree 
%of freedom because ${\bf A}$ is 3-dimensional and two degrees of freedom are used in the fit ($R_{V/W}$ and the amplitude).}
%\begin{tabular}{cclcccccccccc}
%\hline\hline
%Sample & & $\chi^2_{\rm min}$ & & \multicolumn{9}{c}{\mbox{$R_{V/W}$}} \\
%  & & & & $-2\sigma$ & & $-1\sigma$ & & best fit & & $+1\sigma$ & & $+2\sigma$ \\
%\hline
%LRG & & 0.0007 & & 0 & & 0 & & 0.61 & & 2.54 & & $\infty$ \\
%QSO & & 0.67 & & 0 & & 0.41 & & 0.65 & & 0.87 & & 1.36 \\
%NVSS & & 0.02 & & 0.24 & & 0.83 & & 1.17 & & 1.70 & & $\infty$ \\
%\hline
%combined & & 0.002 & & 0.30 & & 0.64 & & 0.87 & & 1.11 & & 1.61 \\
%\hline\hline
%\end{tabular}
%\end{table}

\subsection{Galactic foreground}
\label{ss:gfg}

Galactic foreground emission is an important potential source of 
systematic error for CMB experiments, including those aiming to study 
lensing.  It arises locally and thus cannot correlate with LSS, but it can 
correlate with systematic errors in the LRG and quasar maps, such as 
stellar contamination or errors in the extinction correction.  Galactic 
effects can be treated in basically three ways: (i) one could search for 
frequency dependence in the maps; or (ii) one could feed an estimate of the 
Galactic foreground into the lensing estimator and correlate this with the 
real LSS maps.  Of these, (i) is in principle the best if a wide range of frequencies, e.g. the 
whole WMAP range 23--94 GHz, is available.  Unfortunately the 
lower-frequency bands do not resolve the high multipoles so this approach 
has limited applicability, and in any case the error bars associated with 
this approach are large.  Therefore we also use (ii), which gives much tighter constraints.

A model for the Galactic foreground must include several known 
microwave-emitting components.  In the V and W bands, thermal emission 
from dust and free-free radiation from electron-ion collisions are the 
dominant foregrounds.  At the lower WMAP frequencies there is also a soft 
synchrotron component.  In addition, there is an ``anomalous'' component 
that dominates at the lower WMAP frequencies and appears to be significant 
in V band, which may be due to rotational or magnetic dipole emission from 
dust.

We use a model for the Galactic foreground based on the analysis of 
\cite{2004ApJ...614..186F}; this is the same model as was used in the 
first-year WMAP lensing analysis \cite{2004PhRvD..70j3501H}.  This model 
is composed of the ``Model 8'' \cite{1998ApJ...500..525S, 
1999ApJ...524..867F} prediction for thermal dust emission, added to the 
prediction for free-free radiation based on H$\alpha$ line emission 
\cite{2003ApJS..146..407F} with the conversion factor of 
\cite{2003ApJS..148...97B}.  The anomalous microwave emission is included 
by multiplying the thermal dust template by a factor proportional to 
$T_{\rm dust}^2$, which provides a good phenomenological fit 
\cite{2004ApJ...614..186F}.  The soft synchrotron component is negligible 
in V and W bands, so we do not include it.  (In any event, the usual soft 
synchrotron templates used in CMB work, namely the 408 MHz Haslam maps 
\cite{1981A&A...100..209H, 1982A&AS...47....1H}, have a $\sim 50$ arcmin 
beam that filters out the high multipoles used in the lensing analysis.  
Use of this map as a foreground indicator for lensing or any other 
high-$l$ application would give misleading results.)

Since the Galactic emission is not correlated with the CMB, we may estimate the Galactic contamination to the lensing amplitude $\Delta A$
by feeding the Galactic foreground model maps through the lensing pipeline in place of the actual WMAP maps.  We computed this for each of the 
frequency combinations (VV, VW, and WW) and each sample (LRG, quasar, and NVSS); the worst value of the contamination $|\Delta A|$ was $9\times 
10^{-4}$.  This is negligible compared to our error bars and hence we make no attempt to correct for it.

%\begin{table}
%\caption{\label{tab:foreground}The amplitude $A$ of the lensing cross-correlation signal normalized to that predicted for the WMAP cosmology
%obtained using Galactic foreground model maps instead of the actual WMAP maps.  The Galactic foreground model is described in Section~\ref{ss:gfg}.}
%\begin{tabular}{cllllllll}
%\hline\hline
%Bands & & $A$ (LRGs) & & $A$ (QSOs) & & $A$ (NVSS) & & $A$ (combined) \\
%\hline
%TT & & $+0.00031$ & & $+0.00020$ & & $+0.00065$ & & $+0.00048$ \\
%VV & & $+0.00028$ & & $+0.00041$ & & $+0.00047$ & & $+0.00042$ \\
%VW & & $+0.00033$ & & $+0.00018$ & & $+0.00065$ & & $+0.00048$ \\
%WW & & $+0.00032$ & & $-0.00005$ & & $+0.00087$ & & $+0.00054$ \\
%\hline\hline
%\end{tabular}
%\end{table}

\section{Extragalactic foregrounds}
\label{sec:exf}

This section considers the major extragalactic foregrounds that can contaminate the lensing signal: point sources; thermal SZ fluctuations; kinetic SZ; 
and the Rees-Sciama effect.  The point sources can be subdivided into ``radio'' sources, which emit via synchrotron and free-free radiation and are 
brightest (relative to the CMB) at low frequency, and ``infrared'' sources, i.e. dusty galaxies.  We rely mostly on a combination of frequency and 
spatial dependence of the foregrounds to separate them from lensing.  The brightest radio sources are usually active galactic nuclei (AGN), although 
star-forming galaxies also emit synchrotron and free-free.  Note that it is possible for the same object to be both a radio and an infrared source.

This section is organized as follows.  The spatial dependence of foregrounds is discussed in Section~\ref{ss:ang}, where we show that the bias in the 
lensing signal depends on the galaxy-foreground-foreground bispectrum.  This bispectrum is decomposed using the halo model and we show how to constrain 
each term using the galaxy-temperature cross-spectrum $C_l^{gT}$ and physical constraints on the point sources such as flux cuts.  Most of the rest of 
the section is based on this framework.  We apply the constraints from Section~\ref{ss:ang} to radio point sources in Section~\ref{ss:ps1} and infrared 
sources in Section~\ref{ss:ps2}.  We then investigate the tSZ effect (Section~\ref{ss:tsz}). Several small effects are considered in the appendices: 
Appendix~\ref{app:efg} considers the kSZ and ISW/Rees-Sciama effects, and Appendix~\ref{app:corr} considers correlations between different foreground 
components.  The arguments used to constrain the extragalactic foregrounds are summarized in Section~\ref{ss:exfs}, where we discuss what we believe 
are the most significant strengths and weaknesses and what could be done to improve them in the future.

\subsection{Point sources: spatial correlation-based tests}
\label{ss:ang}

Point sources are the most serious foreground for the WMAP lensing 
analysis.  They introduce two major concerns: one is that the artifacts 
they produce in the lensing map can correlate with large-scale structure, 
thereby introducing a spurious cross-correlation signal; and they can add 
power to the lensing map and thus increase the error bars on the cross 
correlation $C_l^{g{\bf v}}$.  Neither effect is taken into account in the 
simulations.  However the comparison of the lensing power spectra in Section~\ref{ss:ps} demonstrates that the point sources do not add significantly 
to 
the power in the ${\bf v}$ map and hence do not increase the errors $\sigma(C_l^{g{\bf v}})$.

We have several methods of constraining the bias in $C_l^{Xv}$ due to point sources: (i) tests based on the frequency dependence of the signal; (ii) 
tests based on spatial correlations; and (iii) tests based on external foreground maps.  Of these, (i) was considered in Section~\ref{ss:frequency}, 
and 
(iii) will be discussed using the IRAS maps of dusty sources in Section~\ref{ss:iras}.  This section will focus on the methodology for (ii), the tests 
using spatial correlations.  This methodology is applied to radio sources in Section~\ref{ss:ps1} and infrared sources in Section~\ref{ss:ps2}.

Since our estimates of $C_l^{g\kappa}$ are obtained by cross-correlating the LSS map with a ``convergence'' map that is quadratic in temperature, it 
follows that any point source contamination must enter via the cross-bispectrum $B_{ll'l''}^{gTT}$ that contains the LSS tracer and two temperatures.  
The amount of contamination to the $C_l^{g\kappa}$ can be written as
\beq
\Delta C_l^{g\kappa} = R_l^{-1} \sum_{l'l''} {\cal F}_{ll'l''} B_{ll'l''}^{gTT}({\rm ps});
\label{eq:dcl-text}
\eeq
see Appendix~\ref{app:cont} for a derivation and Eq.~(\ref{eq:f}) for the definition of the coefficient ${\cal F}_{ll'l''}$.

We will describe the point source contributions to the cross-bispectrum $B_{ll'l''}^{gTT}$ in the context of a halo model.  Note that since most haloes 
are unresolved by WMAP, the haloes are described entirely by the number of ``galaxies'' (LRGS, quasars, NVSS sources) they contain, their microwave 
flux in each band (in thermodynamic $\mu$K sr), and their 2- and 3-point correlation functions.  In most cases we ignore that the haloes are resolved 
since only the nearest haloes are resolved and in any case we have directly tested that an NFW profile leads to a lower contamination to the lensing 
signal than a delta function with the same integrated flux.
Since very little information is available on the halo 
occupation distributions (HODs) for quasars, NVSS sources, and microwave sources, our analysis will be based to the maximum extent possible on general 
properties of the halo model; in particular we will avoid parameterized HODs entirely.

Within the context of the halo model, the point source contributions to the cross-bispectrum $B_{ll'l''}^{gTT}$ can have four contributions:
\newcounter{cb}
\begin{list}{\arabic{cb}. }{\usecounter{cb}}
\item The ``1-halo'' term, which arises when the two factors of temperature come from point sources in the same halo as the galaxy.
\item The ``2a-halo'' term, in which the two factors of temperature come from point sources in the same halo, but the galaxy is in a different 
halo.
\item The ``2b-halo'' term, in which there is one point source in the same halo as the galaxy and the other point source lives in a different halo.
\item The ``3-halo'' term, in which the galaxy and both point sources live in three distinct halos.
\end{list}
Note that there can be no galaxy-point source-primary CMB cross-bispectrum since the primary CMB resides at $z\sim 1100$ whereas the galaxies and point 
sources are at much lower $z$.

The formulas for $B_{ll'l''}^{gTT}$ are simply stated here so that we can concentrate on the physical assumptions required by our argument; derivations 
for the 1- and 2-halo terms can be found in Appendix~\ref{app:bs}, while the 3-halo term is derived by the usual Limber argument (see e.g. 
Ref.~\cite{2001ApJ...548....7C}).  The number of galaxies $N$ and microwave flux $F$ of a halo are of course dependent not only on the intrinsic 
properties of the halo but also on the redshift, since flux declines as the luminosity distance squared (times the $k$-correction) and the selection 
criteria for our samples are redshift-dependent.  For brevity we will not write this dependence explicitly.  The 1- and 2-halo terms make use of the 
maximum unmasked point source flux $F_{\rm max}$ and the maximum number of galaxies per halo $N_{\rm max}$. We use the notation $\rho(F)\,dF$ is the 
comoving 3-dimensional number density of haloes emitting microwave flux between $F$ and $F+dF$ and $\rho(N)$ to denote the comoving 3-dimensional 
number density of haloes with $N$ galaxies.  The 2-dimensional (projected) number densities will be denoted $n_{2D}(F)$ and $n_{2D}(N)$.  We will also 
use the notation
\beq
f(\chi)\equiv \bar n_g^{-1}r^2 \sum_N N\rho(N) b(N) = \frac{dz}{d\chi}f(z)
\label{eq:fchi}
\eeq
for the bias-weighted comoving distance distribution of the galaxies.

We will consider the 1- and 2a-halo terms together, since they are closely related, and then the 2b- and 3-halo terms separately.  We will use two 
separate ways of estimating the 1- and 2a-halo terms.  Method I uses the cross-power spectra $C_l^{gT}$ obtained in Paper I, and crudely speaking the 
basic idea is that for 1- and 2a-halo terms, the conversion factor from power spectrum to bispectrum cannot be greater than the maximum flux $F_{\rm 
max}$.  Method II looks at the bispectrum of the galaxies and two frequency-differenced maps to constrain the $F^2$-weighted correlation function of 
the galaxies and the point sources.  Generally for types of sources that are very faint, i.e. where $F_{\rm max}$ is below the WMAP noise level per 
beam, Method I gives the tighter constraint, whereas for radio sources (which can be extremely bright) Method II is superior.

\subsubsection{1- and 2a-halo terms: General considerations}

The 1-halo contribution to the cross-bispectrum is simply the Poisson bispectrum,
\beq
B_{ll'l''}^{gTT}({\rm 1h}) = \eta_{ll'l''} \int dF \sum_N \frac{NF^2}{\bar n_g} n_{2D}(N,F),
\label{eq:b1}
\eeq
where $F$ is the halo flux, $N$ is the number of galaxies in the halo, $\bar n_g$ is the mean number of galaxies per steradian, and $n_{2D}(N,F)\,dF$ 
is the number of haloes per steradian containing $N$ galaxies and with flux between $F$ and $F+dF$.  The coefficient $\eta_{ll'l''}$ is given by
\beq
\eta_{ll'l''} = \sqrt{\frac{(2l+1)(2l'+1)(2l''+1)}{4\pi}}\threej{l}{l'}{l''}{0}{0}{0},
\label{eq:etadef}
\eeq
and simply encodes the products of spherical harmonics.  Usually Eq.~(\ref{eq:b1}) would be written in terms of 3-dimensional densities and then 
integrated over redshift (or comoving distance $\chi$) using the Limber approximation.  This is useful for theoretical predictions since theory usually 
gives 3-dimensional densities and biases of haloes as a function of mass and redshift, and does not directly give results in projection.  However in 
our case both the desired constraint (on $B_{ll'l''}^{gTT}$) and the given information ($C_l^{XT}$ measured from Paper I) are already projected onto 
the sky, so we can derive our constraints entirely with Eq.~(\ref{eq:b1}) without considering the 3-dimensional distribution of the point sources.

The analogous equation to Eq.~(\ref{eq:b1}) for the 2a-halo term is
\beqa
B_{ll'l''}^{gTT}({\rm 2ah}) &=& \eta_{ll'l''} \int dF \sum_N \frac{NFF'}{\bar n_g}
\nonumber \\ && \times
n_{2D}(N) n_{2D}(F) C_l(N;F),
\label{eq:b2a}
\eeqa
where $n_{2D}(N)$ is the 2-dimensional density (objects per steradian) of haloes with $N$ galaxies; $n_{2D}(F)\,dF$ is the 2-dimensional density of 
haloes emitting flux between $F$ and $F+dF$; and $C_l(N;F)$ is the cross-power spectrum between haloes containing $N$ galaxies and haloes with flux 
$F$.  This cross-power does not include shot noise.  Once again for the sake of {\em ab initio} theoretical predictions it would be most useful 
to write this as a Limber integral, containing 3-dimensional instead of 2-dimensional densities and involving the 3-dimensional cross-spectrum between 
different types of haloes; but for our purposes we can work entirely in the 2-dimensional space.

This implies a combined 1+2a-halo contribution to the cross-bispectrum,
\beqa
B_{ll'l''}^{gTT}(1+{\rm 2ah}) &=& \eta_{ll'l''} \int dF \sum_N \frac{NF^2}{\bar n_g}
 [n_{2D}(N,F)
\nonumber \\ && 
+ n_{2D}(N) n_{2D}(F) C_l(N;F)].
\eeqa
The corresponding bias in the cross-power is obtained from Eq.~(\ref{eq:dcl-text}):
\beqa
\Delta C_l^{g\kappa}(1+{\rm 2ah}) &=& R_l^{-1}
\left(\sum_{l'l''} {\cal F}_{ll'l''}\eta_{ll'l''}\right)
\nonumber \\ && \times
\int dF \sum_N \frac{NF^2}{\bar n_g}
 [n_{2D}(N,F)
\nonumber \\ &&
+n_{2D}(N)n_{2D}(F) C_l(N;F)].
\eeqa
Note that $l'$ and $l''$ enter only in the summation in parentheses, a fact that will prove to be extremely valuable.  This ``compartmentalization'' is 
a direct result of the pointlike nature of the sources under consideration.  It is convenient to define $r_{ps}(l)$ to be the combination in 
parentheses (this is consistent with the definition in Ref.~\cite{2004PhRvD..70j3501H}), so that
\beqa
\Delta C_l^{g\kappa}(1+{\rm 2ah}) &=&
\frac{r_{ps}(l)}{R_l}
\int dF \sum_N \frac{NF^2}{\bar n_g}
[n_{2D}(N,F)
\nonumber \\ &&
+n_{2D}(N)n_{2D}(F) C_l(N;F)].
\label{eq:contam-12a}
\eeqa
The point source responsivity $r_{ps}(l)/R_l$ is negative for $l\le 269$.  This is a measure of the spurious lensing signal that is produced by a 
point source as a function of scale.  Note in particular that if the galaxies and point sources are positively correlated, i.e.
\beq
\sum_N N[n_{2D}(N,F) + n_{2D}(N) n_{2D}(F) C_l(N;F)] \ge 0,
\label{eq:correl}
\eeq
then the bias $\Delta C_l^{g\kappa}$(1+2ah) must have the same sign as $r_{ps}(l)/R_l$.  Note that for $l\le 269$ we have $r_{ps}(l)/R_l$; this 
is simply a statement that the long-wavelength ($l\le 269$) modes of the reconstructed lensing map have a negative-convergence artifact around each 
point source.  Since our fits to the lensing amplitude $A$ use only the $l\le 269$ multipoles, it follows that it is not possible for the observed 
lensing signal to be due to the 1- or 2a-halo point source terms (i.e. artifacts in the ${\bf v}$ map that arise from individual sources rather than 
interference between sources clustered with each other).  It is possible for the amplitude $A$ to be biased downward (at some level it must be) and 
this must be considered in any cosmological analysis.

\subsubsection{1- and 2a-halo terms: Method I}

Our objective here is to set a constraint on the possible contamination, Eq.~(\ref{eq:contam-12a}).  To do this, we consider the cross-spectrum of the 
galaxies and the temperature, $C_l^{gT}$, coming from the point sources.  This is a relatively simple calculation that has only 1- and 2-halo 
contributions; the answer is
\beqa
C_l^{gT}({\rm ps}) &=& \int dF \sum_N \frac{NF}{\bar n_g} [n_{2D}(N,F)
\nonumber \\ &&
+ n_{2D}(N) n_{2D}(F) C_l(N;F)].
\label{eq:pt}
\eeqa
We now note that the integral over $dF$ in Eq.~(\ref{eq:contam-12a}) is identical to $C_l^{gT}({\rm ps})$ except for an additional factor of $F$.  
If we make the assumption that the fluxes are less than some $F_{\rm max}$, and also assume that the integrand is nonnegative (Eq.~\ref{eq:correl}),
it follows that the integral over $dF$ in Eq.~(\ref{eq:contam-12a}) is bounded between 0 and $F_{\rm max}C_l^{gT}({\rm ps})$.  Then we 
can conclude that (i) the contamination $\Delta C_l^{g\kappa}(1+{\rm 2ah})$ has the same sign as $r_{ps}(l)$, and
(ii) the contamination is bounded by
\beq
|\Delta C_l^{g\kappa}(1+{\rm 2ah})|\le \frac{F_{\rm max}}{R_l} \left| r_{ps}(l)\right|C_l^{gT}({\rm ps}).
\label{eq:bound12a}
\eeq
This simple inequality, which allows us to estimate the point source contamination, relies only on the existence of a bound on the point source flux, 
and the requirement of Eq.~(\ref{eq:correl}), i.e. that galaxies and point sources are positively correlated (where the correlation includes the 
Poisson term).  Note that aside from the positivity of the correlation no assumption is made about the halo mass function, bias, environment 
dependence or lack thereof, etc., so we believe this represents a very robust bound.

\subsubsection{1- and 2a-halo terms: Method II}
\label{ss:m2}

We now consider another way to constrain the 1- and 2a-halo terms.  This time, instead of using a maximum flux $F_{\rm max}$, we attempt to directly 
measure the integral in Eq.~(\ref{eq:contam-12a}) using a quantity quadratic in the CMB temperature and linear in the galaxy density.  We do this using 
frequency-differenced maps, which renders our point source analysis completely insensitive to biases produced by lensing of the CMB, and to noise 
produced by the primary CMB anisotropy.  This method is useful only for the radio point sources since for these Method I is not very constraining 
(due to large $F_{\rm max}$).

The basic idea is that we produce two frequency-difference maps $D^{(1)}$ and $D^{(2)}$.  In the simplest version of the method, the map $D^{(1)}$ is 
the difference of the WMAP temperature maps $T(Ka)-T(V)$, both smoothed to the resolution of the Ka-band map.
The map $D^{(2)}$ is similarly constructed from $T(Q)-T(W)$.  We will denote the 
effective (post-smoothing) beam transfer functions $B^{(1)}_l$ and $B^{(2)}_l$.  We then multiply the two maps to produce 
$D(\nhat) = D^{(1)}(\nhat)D^{(2)}(\nhat)$.  The idea is that each point source produces a feature in $D(\nhat)$ whose intensity is proportional to 
$F^2$ and various factors that involve the frequency dependence.  Specifically, if the input map contains a single point source, then the output map 
intensity can be shown to equal
\beq
D_{lm} = K^{(1)}K^{(2)}F_{V,i}^2 \tau_l Y_{lm}^\ast(\nhat_i),
\label{eq:elm}
\eeq
where $\nhat_i$ is the position of the point source, $F_{V,i}$ is its V-band flux (in units of $\mu$K$\,$sr), $\tau_l$ is the transfer function
\beq
\tau_l = \sum_{l'l''} \frac{(2l'+1)(2l''+1)}{4\pi}B^{(1)}_{l'}B^{(2)}_{l''}\threej{l}{l'}{l''}{0}{0}{0}^2,
\eeq
and the spectral-dependent terms are $K^{(1)} = (F_{Ka,i}-F_{V,i})/F_{V,i}$ and similarly for $K^{(2)}$. (This formula follows immediately from the 
multiplication in real space of the two beams.) If the map contains multiple 
point sources then in addition to terms of the form Eq.~(\ref{eq:elm}) there are interference terms involving combinations of two point sources (with 
different sources contributing to $D^{(1)}$ and $D^{(2)}$).  We may then do a 
cross-correlation with the galaxies, $C_l^{gD}$.  The 1+2a halo contributions (i.e. leaving out the interference contribution to $D$) to this are
\beq
C_l^{gD}(1+{\rm 2ah}) = K^{(1)}K^{(2)}\tau_l 
\int dF\, n_{2D}(F)F^2 C_l(g,F),
\eeq
where $C_l(g,F)$ is the angular cross-spectrum between galaxies and haloes of flux $F$.  This can be expanded as
\beqa
C_l^{gD}(1+{\rm 2ah}) &=& K^{(1)}K^{(2)}\tau_l 
\int dF \sum_N \frac{NF^2}{\bar n_g}
[n_{2D}(N,F)
\nonumber \\ &&
+n_{2D}(N)n_{2D}(F) C_l(N;F)],
\eeqa
so that
\beq
\Delta C_l^{g\kappa}(1+{\rm 2ah}) = \frac{r_{ps}(l)}{R_l}
\frac{
C_l^{gD}(1+{\rm 2ah})  }{K^{(1)}K^{(2)}\tau_l}.
\label{eq:bound12a-ii}
\eeq
We use this equation to estimate the 1+2a 
halo point source contamination of $C_l^{g\kappa}$.

In this method it is important to remove the Galactic foreground, particularly since one is working with the low-frequency bands where the Galaxy is 
brightest.  There are two ways to do this.  One is to subtract the Galactic foreground model (Section~\ref{ss:gfg}).  Alternatively,
instead of directly multiplying the maps $D^{(i)}$, one can high-pass filter them, which should remove more of 
the slowly varying Galactic signal.  The latter method also has the advantage that the temperature offsets in the different bands is not needed 
(WMAP does not measure these directly and used a simple model of the Galaxy to set these offsets).
A simple, local high-pass filter is to subtract the map smoothed by a Gaussian of width $\sigma_1=45\,$arcmin; this is equivalent 
to modifying the effective beam in accordance with
\beq
B^{(i)} \rightarrow B^{(i)}[1-e^{-l(l+1)\sigma_1^2/2}].
\eeq
Thus $\tau_l$ is different for the two methods.

\subsubsection{2b-halo term}

The analogous equation to Eq.~(\ref{eq:b1}) for the 2b-halo term is
\beqa\!\!\!\!
B_{ll'l''}^{gTT}({\rm 2bh}) &=& \eta_{ll'l''} \int dF \int dF' \sum_N \frac{NFF'}{\bar n_g}
\nonumber \\ && \times
n_{2D}(N,F) n_{2D}(F')
\nonumber \\ && \times
[C_{l'}(N,F;F')+C_{l''}(N,F;F')],
\label{eq:b2b}
\eeqa
where $n_{2D}(N)$ is the 2-dimensional density (objects per steradian) of haloes with $N$ galaxies; $n_{2D}(F)\,dF$ is the 2-dimensional density of
haloes emitting flux between $F$ and $F+dF$; and $C_l(N,F;F')$ is the cross-power spectrum between haloes containing $N$ galaxies with flux $F$ and
haloes with flux $F'$.  The spurious contribution to the galaxy-convergence correlation is obtained from Eq.~(\ref{eq:dcl-text}):
\beqa
\Delta C_l^{g\kappa}({\rm 2bh}) &=&
2 R_l^{-1} \sum_{l'l''} {\cal F}_{ll'l''} \eta_{ll'l''}
\nonumber \\ && \times
\int dF \int dF' \sum_N \frac{NFF'}{\bar n_g}
\nonumber \\ && \times
n_{2D}(N,F) n_{2D}(F') C_{l'}(N,F;F').\;\;\;\;
\label{eq:contam-2b}
\eeqa
Here we have removed the $C_{l''}(N,F;F')$ term and doubled the result because of the symmetry of ${\cal F}$ and $\eta$ under interchange of $l'$ and 
$l''$.

We would like to place a bound on $\Delta C_l^{g\kappa}({\rm 2bh})$ using the galaxy-temperature cross power spectrum, just as was done for the 1- and 
2a-halo terms in Method I.  There are two simple ways of doing this: one based on the maximum flux $F_{\rm max}$, and the other based on the maximum 
number of galaxies per halo $N_{\rm max}$.  The former is model-independent and is preferred, except for the radio sources for which it does not give 
an interesting constraint due to large $F_{\rm max}$.

The $F_{\rm max}$ method works by using the triangle inequality in Eq.~(\ref{eq:contam-2b}), and then noting that $F\le F_{\rm max}$:
\beqa
|\Delta C_l^{g\kappa}({\rm 2bh})| &\le&
2 R_l^{-1} \sum_{l'}\left|\sum_{l''} {\cal F}_{ll'l''} \eta_{ll'l''}\right|
\nonumber \\ && \times
\int dF' \sum_N \frac{NF_{\rm max}F'}{\bar n_g}
\nonumber \\ && \times
n_{2D}(N) n_{2D}(F') C_{l'}(N;F').
\eeqa
The last two lines in this equation are -- aside from the factor of $F_{\rm max}$ -- equal to the 2-halo galaxy-temperature spectrum.  This must be 
less than or equal to the actual galaxy-temperature spectrum since the 1-halo term must be nonnegative, so we may write
\beqa
|\Delta C_l^{g\kappa}({\rm 2bh})| &\le&
2 R_l^{-1} \sum_{l'}\left|\sum_{l''} {\cal F}_{ll'l''} \eta_{ll'l''}\right|
\nonumber \\ && \times  F_{\rm max}C_{l'}^{gT}({\rm ps}).
\label{eq:bound2bf}
\eeqa
The last cross-spectrum was directly measured out to $l'=600$.  It is extrapolated to higher $l'$ assuming $C_{l'}^{gT}=$constant, which is 
conservative since the 2-halo $C_{l'}$ decreases at high multipole [due to the declining $P(k)$ over the relevant range of scales].

The model-dependent $N_{\rm max}$ method works as follows.  Written 
in terms of angular correlations, the 2-halo contribution to the point source power spectrum is
\beqa
C_l^{TT}({\rm ps,2h}) &=& \int dF \int dF'\, FF'
n_{2D}(F)
\nonumber \\ && \times
n_{2D}(F') C_{l'}(F;F').
\eeqa
We then have
\beqa
|\Delta C_l^{g\kappa}({\rm 2bh})| &\le &
2 R_l^{-1} \sum_{l'} \left|\sum_{l''}{\cal F}_{ll'l''} \eta_{ll'l''}\right|
\nonumber \\ && \times
\frac{N_{\rm max}}{\bar n_g} C_{l'}^{TT}({\rm ps,2h}),
\label{eq:bound2b}
\eeqa
where $N_{\rm max}$ is the maximum number of galaxies found in a halo and we have assumed the positivity condition
\beq
\sum_N NC_l(N,F;F')\ge 0.
\eeq
On the other hand, written in terms of a Limber integral, $C_l^{TT}$(ps,2h) is
\beq
C_l^{TT}({\rm ps,2h}) = \int \frac{d\chi}{r^2} [b_T\bar T'(\chi)]^2 P_{\rm lin}(k)
\eeq
where $P_{\rm lin}(k)$ is the linear power spectrum at $k=(l+1/2)/r$,
\beq
\bar T'(\chi) \equiv r^2 \int dF\, F\rho(F)
\eeq
is the mean contribution to the brightness temperature per unit comoving distance, and
\beq
b_T \equiv \frac{\int dF\, F\rho(F)b(F)}{\int dF\, F\rho(F)}
\eeq
is the effective (flux-weighted) bias of the microwave emitters.
We note that $\bar T'(\chi)$ does not diverge at small distances even though the fluxes from 
certain haloes can become large because of the $r^2$ factor; this can be seen explicitly from the relation
\beq
\bar T'(\chi,\nu) = \frac{c^2}{2k_{\rm B}\nu^2}(1+z)\bar j^{\rm com}_{\nu(1+z)},
\label{eq:bart-j}
\eeq
where $\nu$ is the observed frequency, $\bar j^{\rm com}_\nu$ is the comoving emissivity (in e.g. erg$\,$cm$^{-3}\,$s$^{-1}\,$Hz$^{-1}\,$sr$^{-1}$, 
where the cm are comoving and all other units are physical), and $k_{\rm B}$ is Boltzmann's constant.

This equation is not quite in a usable form, since the intensity and redshift distribution ($\bar T'$) and bias ($b_T$) of the non-primordial
background are still unknown.  Therefore 
we have tried a different approach.  Using the Limber approximation, and assuming linear
biasing, the second (2-halo) term of Eq.~(\ref{eq:pt}) can be rewritten as
\beqa
\!\!\!\!
C_l^{gT} &\ge& C_l^{gT}({\rm 2h})
\nonumber \\
&=& \int d\chi\,r^2 \int dF \sum_N \frac{NF}{\bar n_g}\rho(F)\rho(N)
\nonumber \\ && \times b(F)b(N)P_{\rm lin}(k),
\nonumber \\
&=& \int \frac{d\chi}{r^2} f(\chi) b_T\bar T'(\chi)P_{\rm lin}(k).
\eeqa
Physically we expect the bias $b_T$ to vary slowly with redshift.  The same is also true of $\bar j_\nu$ and hence $\bar T'(\chi)$ (see 
Eq.~\ref{eq:bart-j}).  If over the range of redshifts of interest it is a power law, $b_T\bar T'(\chi)\propto (1+z)^\zeta$, then
\beq
b_T\bar T'(\chi)|_{z=0} \le \frac{C_{l_0}^{XT}}{\int d\chi\, r^{-2} f(\chi) (1+z)^\zeta P_{\rm lin}(k_0)},
\label{eq:btbt}
\eeq
where equality would hold if the right-hand side included only the two-halo term, $l_0$ is any arbitrary multipole, and $k_0=(l_0+1/2)/r$.  In practice 
to increase signal-to-noise ratio, we average the right-hand side over all the multipole bins between 25 and $l_{\rm max}$ (the multipole 
corresponding to $k=0.1h\,$Mpc$^{-1}$ at the 20th percentile in distance).  We construct bounds on the 2b-halo 
point source contamination by considering a range of exponents $\zeta$.  Of course, all of these integrals to the redshift range where there are 
galaxies; we use $z_{\rm max}=0.8$ for the LRGs, 2.7 for the quasars, and 3 for NVSS.

For the quasars and the NVSS galaxies the constraint on the maximum number of galaxies per halo will be redshift-dependent.  In this case, 
it is straightforward to show that Eq.~(\ref{eq:bound2b}) undergoes the replacement
\beq
N_{\rm max}C_{l'}^{TT}({\rm ps,2h}) \rightarrow
\int \frac{d\chi}{r^2} N_{\rm max}(\chi)[b_T\bar T'(\chi)]^2 P_{\rm lin}(k),
\eeq
i.e. $N_{\rm max}$ is pulled inside the integral over comoving distance (or redshift).

\subsubsection{3-halo term}

The 3-halo contribution to the point source bispectrum is different from the 1- and 2-halo terms in that we have found no simple model-independent 
argument to constrain 
it using the galaxy-temperature correlation.  Instead we will have to use a more model-dependent method.  The Limber equation allows us to express the 
galaxy-temperature-temperature angular bispectrum in terms of the 3-dimensional bispectrum,
\beqa
B_{ll'l''}^{gTT}({\rm 3h}) &=& \eta_{ll'l''} \int d\chi\,r^2 \int dF \int dF' \sum_N \rho(N)
\rho(F)
\nonumber \\
&& \times
\rho(F')
 \frac{NFF'}{\bar n_X} B(k,k',k''|N;F;F'),
\label{eq:b-ang}
\eeqa
where $\chi$ is the radial comoving distance; $r$ is the comoving angular diameter distance (equal to $\chi$ for a flat universe);
$\rho(N)$ is the comoving number density of haloes containing $N$ galaxies; $\rho(F)dF$ is the comoving number density of haloes with microwave 
flux between $F$ and $F+dF$; the wavenumbers are $k=(l+1/2)/r$, etc.; and $B(k,k',k''|N;F;F')$ is the 3-dimensional cross-bispectrum of the haloes 
with $N$ galaxies, the haloes with flux $F$, and the haloes with flux $F'$.  The problem is that the key quantities $\rho(F)$ and 
$B(k,k',k''|N;F;F')$ are not known.

To circumvent this problem, we will consider the linear bias prediction for the 3-halo term.  That is, we set
\beq
B(k,k',k''|N;F;F') = b(N)b(F)b(F')B_{\rm 2P}(k,k',k''),
\label{eq:3x}
\eeq
where $b$ represents the bias of the haloes and $B_{\rm 2P}(k,k',k'')$ is the second-order perturbation theory bispectrum.  This allows us to factor 
Eq.~(\ref{eq:b-ang}), yielding
\beq
B_{ll'l''}^{gTT} = \eta_{ll'l''} \int \frac{d\chi}{r^4} f(\chi) [b_T\bar T'(\chi)]^2
B_{\rm 2P}(k,k',k'').
\eeq
Using Eq.~(\ref{eq:dcl-text}), we may then write
\beqa
|\Delta C_l^{g\kappa}({\rm 3h})| &\le& R_l^{-1}\sum_{l'l''}\Bigl|{\cal F}_{ll'l''}\eta_{ll'l''}
\int \frac{d\chi}{r^4} 
\nonumber \\ && \times
f(\chi) [b_T\bar T'(\chi)]^2
B_{\rm 2P}(k,k',k'')\Bigr|.
\label{eq:bound3}
\eeqa
The factor $b_T\bar T'(\chi)$ comes from the bound, Eq.~(\ref{eq:btbt}), just as for the 2b-halo term ($N_{\rm max}$ method).
Like the 2b-halo term, it is model-dependent.  Fortunately, the bound is also very small so deviations of halo clustering 
from the simple perturbation theory predictions or more complicated redshift evolution of $b_T\bar T'(\chi)$ will have little effect.

\subsection{Estimates of contamination: radio sources}
\label{ss:ps1}

We now construct radio point source contamination estimates using the methodology of the previous section.  For the 2b- and 3-halo terms we need to 
know the contribution to the cross-spectra $C_l^{gT}$ from radio sources in order to get $b_T\bar T'$.  We will use only Method II for the 1+2a-halo 
terms because for the bright radio sources Method I does not give an interesting constraint.

The $C_l^{gT}$ cross-powers for the LRGs, the SDSS quasars, and the NVSS sources were obtained in 
Paper I in the range up to $l=600$.  Unfortunately they are noisy, with the noise dominated by primary CMB fluctuations, and on large angular scales 
the ISW effect dominates the cross-power.  For this reason, we have used the differences $C_l^{gT}(Q)-C_l^{gT}(V)$ and $C_l^{gT}(V)-C_l^{gT}(W)$ as the 
major sources of information on point sources.  We construct our estimates for the V-band lensing estimator, which will be the most contaminated by 
radio sources.  For flat spectrum sources and the relative weighting of bands in Eq.~(\ref{eq:combo}), the ratio of contamination is
\beq
\frac{\Delta C_l^{g\kappa}(TT)}{\Delta C_l^{g\kappa}(VV)} = 0.56.
\eeq

We take as our estimate of the radio point source cross-spectrum
\beqa
\hat C_l^{gT}(V;{\rm rps}) &\equiv& a_1[C_l^{gT}(Q)-C_l^{gT}(V)]
\nonumber \\ &&
 + a_2[C_l^{gT}(V)-C_l^{gT}(W)],
\label{eq:rps1}
\eeqa
with $a_1 = 1.128$ and $a_2 = -0.466$.  The ratio $a_2/a_1=-0.413$ is chosen to cancel any tSZ contribution,
and the normalization was chosen to be correct for flat-spectrum radio sources with 
$F_\nu=$constant.  If there are also steep-spectrum sources this is an overestimate of their effect: for $F_\nu\propto \nu^{-0.8}$ we have $\hat 
C_l^{gT}(V;{\rm rps}) = 1.66C_l^{gT}(V;{\rm rps})$.  Equation~(\ref{eq:rps1}) has some sensitivity to infrared sources: for 
$F_\nu\propto\nu^{3.5}$ we have $\hat C_l^{gT}(V;{\rm rps}) = 0.008C_l^{gT}(V;{\rm dust})$, and the coefficient 0.008 rises to 0.146 if 
$F_\nu\propto\nu^4$.  Since this coefficient is positive we will overestimate the contribution of radio sources if the infrared sources are also 
significant.  This cross-spectrum is obtained by constructing a WMAP map of the frequency combination 1.128Q-1.594V+0.466W and cross-correlating it 
with each galaxy map.  Also in order to speed up the calculation we did
not implement any weighting in the $C_l^{gT}$ estimation, i.e. instead of full ${\bf C}^{-1}$ weighting we only project out the monopole and dipole
from each map (and in the case of NVSS, the declination rings and Haslam map).  We have computed error bars for each
case by cross-correlating 20 {\em simulated} LSS maps with the actual WMAP difference map (on large angular scales in WMAP the difference maps 
contain $1/f$ noise and Galactic foregrounds, and hence are more difficult to reliably simulate than the LSS maps).  The simulated LSS maps are 
Gaussian and have the correct autopower spectra as determined from Paper I.  The Gaussianity of the LSS does not matter for getting the variance of 
$C_l^{gT}(V;{\rm rps})$, since the $C_l^{gT}(V;{\rm rps})$ estimator is a linear function of the data.

The 2b- and 3-halo terms can be constrained if we know the maximum number of galaxies per halo $N_{\rm max}$ (needed for the 2b-halo term only) and the 
redshift evolution exponent $\zeta = d\ln[b_T\bar T'(\chi)]/d\ln(1+z)$.  We estimate $N_{\rm max}$ for NVSS by examining each object and finding the 
number of neighbors (including the object itself) within 6 arcminutes ($\sim 1h^{-1}\,$Mpc at $z=0.2$).  For NVSS there are no objects with $>12$ 
neighbors, 1 with 12 neighbors, and 2 with 11 neighbors, so we set $N_{\rm max}=12$.  This may be a large overestimate: in particular Poisson 
statistics predicts 0.1 objects with 12 neighbors and 0.9 with 11 neighbors, so it is not clear that these groupings should be taken seriously.  
Nevertheless for our purposes all we need is an upper limit, so we use 12.  For the quasars the maximum number of neighbors is 10, so we set $N_{\rm 
max}=10$.  Again this should not be taken especially seriously (for Poisson statistics there should be on average 1.2 objects with 10 neighbors).  For 
the LRGs a similar argument gives $N_{\rm max}=36$.  For quasars and NVSS the redshift distribution extends to higher $z$ where we can improve the 
constraint on $N_{\rm max}$.  For example at $z=0.8$ a $1h^{-1}\,$Mpc radius corresponds to 2 arcminutes; thus for $z\ge 0.8$ we have replaced the 
above $N_{\rm max}$ values with the maximum number of neighbors within 2 arcmin.  This is 5 for the quasars and 5 for NVSS.  For NVSS we have also set 
$N_{\rm max}=3$ for $z\ge 3$, which is what we find with a 45 arcsec radius ($1h^{-1}\,$Mpc at $z=3$).

We do not know the redshift dependence of $b_T\bar T'$ so we have to consider a range of possibilities and determine which leads to the most serious 
contamination.  This is done for power laws $b_T\bar T'\propto (1+z)^\zeta$ in the last 
part of Table~\ref{tab:rps} (note that flat spectrum sources with constant comoving density would have $\zeta=2$).  Note that the 2b- and 3-halo terms 
are shown to be negligible.  For $\zeta=0$ we find upper limits to $b_T\bar 
T'(\chi)$ to be $(0\pm 5)h\,$nK$\,$Mpc$^{-1}$ for the LRGs, $(-6\pm 11)h\,$nK$\,$Mpc$^{-1}$ for the quasars, and $(+1\pm 2)h\,$nK$\,$Mpc$^{-1}$ for 
NVSS; these are upper limits because they assume the entire correlation comes from the 2-halo term.  The combined 95\%\ confidence upper limit is 
$4h\,$nK$\,$Mpc$^{-1}$.  This is a one-tailed upper limit ($+1.64\sigma$) since physially $b_T\bar T'$ should be positive.

We now apply Method II (Section~\ref{ss:m2}) to the 1+2a halo term.  For the method based on $Ka-V$ and $Q-W$ differences, and for flat-spectrum 
sources the flux ratios $(F_{Ka}-F_{V})/F_{V}$ and $(F_{Q}-F_{W})/F_{V}$ are 2.20 and 1.63, respectively, so that $K^{(1)}K^{(2)}=3.57$.  The transfer 
function $T_l$ varies from $5.0\times 10^3$ to $3.0\times 10^3$ as $l$ varies from 0 to 400.  The implied bias $\Delta A$ is given in Table~\ref{tab:rps}.  We have shown this for both cases where the Galactic foreground model is 
subtracted (``Gs'') and where it is removed by high-pass filtering (``hpf'').  We have also shown this for different combinations of frequency bands.  
In cases where the same band appears twice, e.g. (Ka$-$V)(Q$-$V), we have computed only cross-correlations involving different DAs and averaged the 
results, e.g. ``(Ka$-$V)(Q$-$V)'' is actually the average of (Ka$-$V1)(Q$-$V2) and (Ka$-$V2)(Q$-$V1).  Just as for Method I we compute error bars using 
simulated LSS maps since it is harder to simulate the product-difference maps.

\begin{table*}
\caption{\label{tab:rps}Estimates of the radio point source contamination $|\Delta A|$ to the galaxy-convergence correlation 
amplitude $A$.  The estimates are referenced to the VV lensing maps; for flat-spectrum sources the contamination to TT is reduced by a factor of 0.56.
We consider separately the contribution from the 1+2a, 2b, and 3-halo terms.  The first part of the table consists of upper limits; the subscript 
``max'' is used to emphasize that these are estimates of an upper limit.  In particular $|\Delta A|_{\rm max}$ can be used to assess the magnitude of 
systematic errors but should not be used to ``correct'' the data!  The bottom part of the table contains estimates of the contamination -- these are 
``central'' values and their errors (where shown) are $1\sigma$ statistical errors, not including any uncertainty in the frequency dependence of the 
foreground.  The notations ``Gs'' and ``hpf'' are used to indicate 
that the Galactic foreground is subtracted (Gs) from the difference maps $D^{(1,2)}$, or that it is removed by high-pass filtering (hpf); see 
Section~\ref{ss:m2}.}
\begin{tabular}{lllllllll}
\hline\hline
Term & ~~ & Method/assumption & ~~ & $|\Delta A|_{\rm max}$(LRG) & ~~ & $|\Delta A_{\rm max}|$(QSO) & ~~ & $|\Delta A|_{\rm max}$(NVSS) \\
\hline
2b halo & & $\zeta=+4$, fiducial cosmology & & $8.8\times 10^{-4}$ & & $2.8\times 10^{-2}$ & & $4.0\times 10^{-2}$ \\
2b halo & & $\zeta=0$, fiducial cosmology & & $5.6\times 10^{-3}$ & & $6.2\times 10^{-3}$ & & $7.0\times 10^{-3}$ \\
2b halo & & $\zeta=-4$, fiducial cosmology & & $1.8\times 10^{-2}$ & & $1.6\times 10^{-2}$ & & $1.9\times 10^{-2}$ \\
\hline
3  halo & & $\zeta=0$, fiducial cosmology & & $1.2\times 10^{-3}$ & & $1.4\times 10^{-4}$ & & $4.3\times 10^{-4}$ \\
3  halo & & $\zeta=+4$, fiducial cosmology & & $1.7\times 10^{-4}$ & & $6.5\times 10^{-4}$ & & $4.2\times 10^{-4}$ \\
3  halo & & $\zeta=-4$, fiducial cosmology & & $1.7\times 10^{-3}$ & & $1.4\times 10^{-4}$ & & $1.1\times 10^{-3}$ \\
\hline
Term & ~~ & Method/assumption & ~~ & $\Delta A$(LRG) & ~~ & $\Delta A$(QSO) & ~~ & $\Delta A$(NVSS) \\
\hline
1+2a halo & & Method II, (Ka$-$V)(Q$-$W), hpf       & & $+0.060\pm0.045$ & & $+0.039\pm0.040$ & & $+0.041\pm0.037$ \\
1+2a halo & & Method II, (Ka$-$V)(Q$-$W), Gs        & & $+0.001\pm0.076$ & & $-0.007\pm0.062$ & & $-0.023\pm0.034$ \\
1+2a halo & & Method II, (Ka$-$V)(Q$-$V), hpf       & & $+0.005$ & & $-0.013$ & & $+0.046$ \\
1+2a halo & & Method II, (K$-$Q)(Ka$-$V), hpf       & & $+0.032$ & & $+0.013$ & & $+0.021$ \\
1+2a halo & & Method II, (Ka$-$V)(Q$-$V), Gs        & & $-0.090$ & & $-0.009$ & & $-0.026$ \\
\hline\hline
\end{tabular}
\end{table*}

\subsection{Estimates of contamination: infrared sources}
\label{ss:ps2}

Infrared sources present a different challenge than radio sources.  On the one hand, our direct constraints on $F_{\rm max}$ are quite weak due to the 
lack of coverage at frequencies above W band: there are no all-sky maps at frequencies between the WMAP W band (94 GHz) and COBE/DIRBE channel 10 (1250 
GHz).  However most infrared sources are quite weak (there are no sources in the WMAP catalog with spectral indices consistent with thermal dust 
emission) so taking $F_{\rm max}$ from the WMAP detection threshold would be very conservative (in fact, it would not yield a useful constraint).  
Therefore Method I {\em by itself} is not very helpful.  On the other hand, the same low flux of the point sources makes Method II useless as any 
frequency-difference maps are dominated by Galactic emission on large scales and radio sources on small scales.

In the face of these difficulties, we have resorted to a combination of two different approaches.  The key idea behind our approach is that infrared 
sources at cosmological distances have fluxes far below the WMAP detectability threshold.  (Note that this is very different from the behavior of radio 
AGNs.)  Therefore we consider separately the distant sources ($z\ge 0.02$) and nearby sources ($z<0.02$).  The distant sources are constrained using 
the arguments of Section~\ref{ss:ang}, including Method I for the 1+2a-halo terms.  The redshift cut allows us to use a small value of $F_{\rm max}$ 
(essentially the flux of an object a few times $L_\star$ at $z=0.02$).  The nearby infrared sources can be handled by a very different argument: they 
appear in the IRAS 100$\,\mu$m maps, so we can use these maps to constrain them, i.e. we can feed the IRAS maps through our lensing pipeline with an 
appropriate scaling factor.  (Distant infrared sources at $z\sim 1$, which are a major source of concern for us, may appear in WMAP but not IRAS 
because the $k$-correction for a thermal dust spectrum acts to brighten these sources in V and W bands, but make them fainter in the IRAS bands.)  
Being at different redshifts, the two groups of sources cannot have any physical correlation with each other.  The sources at $z<0.02$ should have very 
little overlap with our large-scale structure samples, especially the LRGs, but there are a few quasars with low spectroscopic redshifts, and NVSS must 
have a significant tail at $z<0.1$ because of the nonzero NVSS$\times$2MASS cross-correlations.  Therefore these low-redshift sources must be 
constrained in order to have a reliable result.

Note that many dusty sources at $z>0.02$ will appear in the IRAS maps (in the sense that the IRAS maps contain their statistical fluctuations, even if 
the sources themselves are confused or buried under Galactic emission).  This means that these sources are double-counted in our argument, once in the 
statistical investigation of $C_l^{gT}$ and once in the foreground tests using IRAS maps.  This does not represent a problem since it only serves to
make our bounds more conservative.

\subsubsection{Sources at $z\ge 0.02$}

We now apply the methods of Section~\ref{ss:ang} to the infrared sources.  The infrared sources are subdominant contributions to the power spectrum in 
all 
of the WMAP bands, so we cannot easily implement Method II.  On the other hand, they are much fainter than the radio sources, so Method I (based on 
$F_{\rm max}$) is extremely useful.  We will reference our constraints on their contamination to the W band since infrared sources are brightest there.  
For $F_\nu\propto\nu^{3.5}$ spectra and the relative weighting of bands in Eq.~(\ref{eq:combo}), the ratio of contamination is
\beq
\frac{\Delta C_l^{g\kappa}(TT)}{\Delta C_l^{g\kappa}(WW)} = 0.51;
\eeq
for $F_\nu\propto\nu^4$ the coefficient is 0.44.

Just as we did for the radio sources, we may construct a linear combination of differences $C_l^{gT}(Q)-C_l^{gT}(V)$ and $C_l^{gT}(V)-C_l^{gT}(W)$ to 
estimate $C_l^{gT}(W)$ without any contamination by the ISW effect.  The difference between V and W bands is the most sensitive to infrared sources, 
but it contains an opposite contribution from radio sources, which must be canceled by involving the Q band.  The appropriate combination is
\beqa
\hat C_l^{gT}(W;{\rm irps}) &\equiv& a_3[C_l^{gT}(Q)-C_l^{gT}(V)]
\nonumber \\ &&
 + a_4[C_l^{gT}(V)-C_l^{gT}(W)],
\label{eq:irps1}
\eeqa
where $a_3=+1.086$ and $a_4=-2.295$.  This combination has no sensitivity to flat-spectrum sources and has the correct normalization for 
sources with spectrum $F_\nu\propto\nu^{3.5}$.  If the spectrum of the infrared source is steeper than $\nu^{3.5}$ (e.g. $\nu^4$) then this equation 
overestimates $C_l^{gT}(W;{\rm irps})$.  If there is a positive contribution to $C_l^{gT}$ from free-free or steep-spectrum sources 
($F_\nu\propto\nu^\alpha$ with $\alpha<0$) then it adds a positive contribution to the estimate $\hat C_l^{gT}(W;{\rm irps})$.  Also it is easily 
verified that for a tSZ spectrum with blackbody temperatures in the ratio $T_Q:T_V:T_W=0.957:0.906:0.783$ and negative amplitude ($C_l^{gT}<0$ as 
appropriate since tSZ produces a decrement in high-density regions), a positive bias is produced in $\hat C_l^{gT}(W;{\rm irps})$.  Therefore 
Eq.~(\ref{eq:irps1}) represents an upper limit on the contribution to $C_l^{gT}(W)$ from infrared sources.

We use Method I here for the 1+2a-halo terms, so we need an estimate of $F_{\rm max}$.  For the radio sources we used the detection threshold in the 
WMAP point source 
catalog, however this limit is too weak to be useful for constraining infrared sources because WMAP does not have any bands at higher frequency than W 
and consequently the sensitivity to infrared sources is quite weak.  (Indeed, the WMAP point source catalog \cite{2007ApJS..170..288H} contains no objects 
consistent with being infrared rather than radio sources.)  Therefore we take a different approach.  The brightest sources (in Jy) may be either local 
objects (bright because they are nearby) or high-redshift submillimeter galaxies.  For the local objects we use the local SCUBA luminosity function at 
353$\,$GHz (the lowest measured frequency) from Dunne et~al. \cite{2000MNRAS.315..115D}, which implies that within our survey solid angle and at $z\ge 
0.02$ there should be on average 0.05 objects with $F>0.5\,$Jy.  Thus we take $F_{\rm max}=0.5\,$Jy at 353$\,$GHz, which is exceeded 
with 5\%\ probability.
High-redshift submillimeter galaxies have typical fluxes of up to tens of mJy at 353$\,$GHz \cite{2006MNRAS.370.1057S, 2006MNRAS.372.1621C}
and hence do not increase our above estimate of $F_{\rm max}$.
In order to be useful, this estimate of the maximum luminosity of the infrared sources must be extrapolated down to W band.  Unfortunately there is 
very little data to suggest the correct form of the SED in this frequency range.  This depends on the value of the emissivity exponent $\beta$ for the 
infrared sources (recall that $F_\nu\propto\nu^{2+\beta}$ in the Rayleigh-Jeans limit).  While low values of $\beta\sim
1.3$ are derived by Dunne et~al. \cite{2000MNRAS.315..115D} from the SED fits for local galaxies (recall that $F_\nu\propto\nu^{2+\beta}$ in the
Rayleigh-Jeans limit), Dunne et~al. also note that these values could be biased low by a superposition of several dust temperatures, and that fits to
our own Galactic emission give larger $\beta$.  We have assumed $\beta=1.3$ to be conservative, which gives $F_{\rm max}=0.013\,$Jy in W band (this 
number decreases if $\beta$ is increased).  In blackbody temperature units this is $F_{\rm max}=0.06\,$nK$\,$sr.

The constraints we obtain are shown in Table~\ref{tab:irps}.  The 1+2a and 2b-halo terms are based on the 95\% confidence upper limit to $-\Delta A$ 
with the specified value of $F_{\rm max}$ (since $\Delta A$ is negative).

For the 3-halo terms we need $b_T\bar T'$.  Unfortunately in this case the galaxy-temperature spectrum is not good enough to constrain $b_T\bar T'$ 
because for infrared sources $b_T\bar T'$ is probably a strongly increasing function of redshift.  This can be seen from Eq.~(\ref{eq:bart-j}): if the 
comoving density of infrared emitters is constant, then due to the $k$-correction we have $\bar T'\propto (1+z)^{1+\alpha}$ where the spectral index 
$\alpha$ is typically $\sim 3.5$.  Also the bias is usually higher for high-redshift galaxies where the fluctuation amplitude is less and hence 
galaxies can only form in rare, highly biased peaks of the density field.  Finally the star formation rate and hence (probably) the infrared luminosity 
density declines at $z<1$ (e.g. \cite{2006ApJ...653..881N}).  Given the very large uncertainties in the redshift distribution of the far-IR background, 
we have tried several models for $\bar T'$.  In Model A, we assume that at $1<z<5$ the comoving infrared emissivity is constant and $\alpha=3.5$ so 
that $\bar T'\propto(1+z)^{4.5}$; at $z<1$ it is assumed that the emissivity declines as $\propto(1+z)^3$ so that $\bar T'\propto(1+z)^{7.5}$.  We 
normalize the model to produce $33\,\mu$K blackbody temperature at W band over the redshift range $z<5$, consistent with extrapolation of the far-IR 
background observed by COBE/FIRAS \cite{1998ApJ...508..123F}.  This normalization may be conservative: there must be some contribution to the far-IR 
background from $z>5$, particularly at low frequencies where the $k$-correction favors high redshifts. The other models include several variations on 
this.  Model B does not include the decline at low redshift, and takes $\bar T'\propto(1+z)^{4.5}$ all the way down to $z=0$; it is designed to 
maximize the contamination for the LRGs.  Models C and D are similar to Model A except that the exponent of $1+z$ is increased or decreased by $\pm 2$ 
respectively. For the bias, we have assumed a constant clustering strength with $\sigma_8(z,{\rm gal})=2$, i.e. $b=2/\sigma_8G(z)$ where $G$ is the 
growth factor. This is somewhat larger than what is observed for the LRGs and quasars, and is probably typical for the most massive objects.  The 
actual emissivity-weighted mean bias is probably somewhat lower.  The resulting estimates of the 3-halo infrared source contamination are shown in 
Table~\ref{tab:irps}.  While the calculation is obviously very crude, it is clear from the table that the infrared source 3-halo term is not a 
significant source of contamination.

\begin{table*}
\caption{\label{tab:irps}Estimates of the infrared point source contamination to the galaxy-convergence correlation 
amplitude $A$.  The estimates are referenced to the WW lensing maps; for $F_\nu\propto\nu^4$ sources the contamination to TT is reduced by a factor of 
0.51.  We consider separately the contribution from the 1+2a, 2b, and 3-halo terms.  The nearby sources are not accounted for in our analysis of the 
1+2a halo terms, so we have used the IRAS maps for these; those estimates are shown in the last line of the table, and are estimates of $\Delta A$, 
not upper limits.}
\begin{tabular}{lllllllll}
\hline\hline
Term & ~~ & Method/assumption & ~~ & $|\Delta A|_{\rm max}$(LRG) & ~~ & $|\Delta A_{\rm max}|$(QSO) & ~~ & $|\Delta A|_{\rm max}$(NVSS) \\
\hline
1+2a halo, $z\ge 0.02$ & & Method I & & $0.0089$ & & $0.0087$ & & $0.0003$ \\
\hline
2b halo & & $F_{\rm max}$ & & $0.0024$ & & $0.0062$ & & $0.0040$ \\
\hline
3  halo & & Model A, fid. cosmology  & & $6.6\times 10^{-7}$ & & $6.0\times 10^{-5}$ & & $1.2\times 10^{-4}$ \\
        & & Model B, fid. cosmology  & & $3.3\times 10^{-6}$ & & $6.0\times 10^{-5}$ & & $1.2\times 10^{-4}$ \\
        & & Model C, fid. cosmology  & & $6.6\times 10^{-9}$ & & $7.3\times 10^{-6}$ & & $9.1\times 10^{-5}$ \\
        & & Model D, fid. cosmology  & & $4.6\times 10^{-5}$ & & $4.4\times 10^{-4}$ & & $3.1\times 10^{-4}$ \\
        & & Model A, high $\sigma_8$ & & $8.7\times 10^{-7}$ & & $6.6\times 10^{-5}$ & & $1.3\times 10^{-4}$ \\
        & & Model A, high $\Omega_m$ & & $1.6\times 10^{-6}$ & & $1.0\times 10^{-4}$ & & $1.8\times 10^{-4}$ \\
\hline
Term & ~~ & Method/assumption & ~~ & $\Delta A$(LRG) & ~~ & $\Delta A$(QSO) & ~~ & $\Delta A$(NVSS) \\
\hline
sources at $z<0.02$ & & Extrapolated from IRAS 100$\mu$m map & & $+2\times 10^{-3}$ & & $+7\times 10^{-5}$ & & $+9\times 10^{-3}$ \\
\hline\hline
\end{tabular}
\end{table*}

\subsubsection{Sources at $z<0.02$: constraints from IRAS}
\label{ss:iras}

The nearby infrared sources are constrained using the $100\,\mu$m maps from the IRAS Sky Survey Atlas \cite{1994issa.book.....I, 
1994STIN...9522539W}, which cover $\sim$98\%\ of the sky with $\sim$6 arcmin resolution.  Our strategy is to feed a rescaled version of the 
IRAS maps through our lensing pipeline, and use the result as an estimate of the contribution to $C_l^{g\kappa}$ from nearby dusty galaxies.  (More 
accurately, we do the rescaling assuming a low dust temperature to be conservative, and treat this as an upper bound on the contamination.)  We note 
that the IRAS maps are dominated by emission from our {\em own} galaxy: this should not represent a problem, as the output of the lensing pipeline 
should then contain the contamination to ${\bi v}$ from both Galactic and nearby-extragalactic dust.

In order to implement the above program, we need to select an SED with which to extrapolate from 100$\,\mu$m to V/W bands, in particular we need the 
ratios $I_\nu(V)/I_\nu(100\mu{\rm m})$ and $I_\nu(W)/I_\nu(100\mu{\rm m})$.  If we assume that the nearby galaxies have SEDs similar to the 
high-latitude regions of the Milky Way, and use the ``cold region'' spectrum from Fig.~1 Finkbeiner et~al. \cite{1999ApJ...524..867F}, we estimate
$I_\nu(W)/I_\nu(100\mu{\rm m})=0.0012$.  The regions of higher dust temperature would have a smaller ratio (bluer spectrum) and hence this is 
conservative.  We mask Saturn from the IRAS maps since planet-contaminated data is not used in WMAP mapmaking.

The resulting contamination estimates in W-band are $\Delta A=+0.002$ for the LRGs, $\Delta A=+7\times 10^{-5}$ for the quasars, and $\Delta A=+0.009$ 
for NVSS.  This is negligible compared to our error bars.

\subsection{Thermal SZ effect}
\label{ss:tsz}

The thermal SZ effect is another possible contaminant of the lensing signal.  This section evaluates contamination from the galaxy-tSZ-tSZ bispectrum.  
The method of setting constraints for the 2b- and 3-halo terms is very similar to the approach for point sources, except that we estimate $b_T\bar T'$ 
from theory rather than from cross-correlation of the galaxies with the temperature field.  This is because on the one hand the weak frequency 
dependence of tSZ (within the WMAP bands) makes the observational measurement of the galaxy-tSZ cross-correlation very noisy; and on the other hand we 
have at least a rough idea of how and why the tSZ decrement depends on halo properties (as opposed to point sources, whose halo occupation properties 
can only be determined empirically).  The 1- and 2a-halo terms can also be constrained theoretically from the mass function and mass-bias relation (for 
the 1-halo term we also need $N_{\rm max}$).  They diverge at low redshift and so we have estimated the contribution from nearby haloes by using their 
actual positions from X-ray data.  The bounds in this section will be written in terms of the Rayleigh-Jeans (RJ) limit of the tSZ effect; since WMAP 
is not quite in the RJ limit, the galaxy-tSZ-tSZ bispectrum is suppressed by a factor of 0.82 (for the VV frequency combination), 0.71 (VW), 0.61 (WW), 
or 0.72 (for TT, the frequency-averaged lensing map).

The nearby haloes are obtained from the Ebeling {\em et~al.} XBAC catalog \cite{1996MNRAS.281..799E} of optical clusters detected in the ROSAT 0.1--2.4 
keV X-ray band.  We have used the X-ray temperature to SZ luminosity conversion normalized according to Hinshaw {\em et~al.} \cite{2007ApJS..170..288H} 
[i.e. we used their Eq.~(29) multiplied by either the central value of the normalization $-0.32$, or the one-sided 95\%\ confidence lower limit 
$-0.55$].  Some of the redshifts in the XBAC catalog are proprietary but there are now publicly available redshifts for these clusters 
\cite{1999ApJ...514..148D, 2002ApJS..139..313D, 2002ApJS..140..239C}, which we use.  With this data it is possible to construct the actual realization 
of the tSZ sky contributed by the nearby X-ray luminous clusters, and to feed this through the lensing pipeline to obtain the contamination $\Delta A$, 
if one knows what assumption to make for the tSZ radial profile of the clusters.  We found earlier that Gaussian or NFW profiles always produce 
smaller lensing contamination than delta functions, so it is conservative to treat the cluster as a delta function.  We do this for all clusters except 
the brightest (Coma); following Hinshaw {\em et~al.} \cite{2007ApJS..170..288H} we have modeled Coma using a $\beta$ profile \cite{1992A&A...259L..31B} 
with central temperature decrement at 94 GHz of either $-0.42\,$mK (best value) or $-0.54\,$mK (95\%CL, one-sided) \cite{1995ApJ...449L...5H}.

The 2b- and 3-halo terms for the tSZ effect depend on the product $b_T\bar T'$ for the tSZ effect.  In the RJ limit, and assuming that the 
intracluster gas contains the cosmic baryon fraction, this is given by
\beq
b_T\bar T'({\rm tSZ}) = -2 \bar\tau'T_{\rm CMB} \int\! \frac{f_{\rm ICM}}{f_b} b(M) \frac{k_{\rm B}T_e(M)}{m_ec^2} \phi(M) dM,
\eeq
where $\bar\tau'$ is the mean optical depth per unit comoving distance, $b(M)$ is the bias for haloes of mass $M$, $k_{\rm B}$ is Boltzmann's constant, 
$m_e$ is the electron mass, $T_e(M)$ is the density-weighted mean electron temperature in haloes of mass $M$, and $\phi(M)\,dM$ is the 
fraction of the mass in haloes of mass between $M$ and $M+dM$.  Here $f_{\rm ICM}/f_b$ is the ratio of the intracluster medium baryon fraction to the 
cosmic value.  This is $\le 1$ in real clusters, since some of the gas has turned into stars or is present in atomic or molecular phases that do not 
contribute to tSZ; however we will set $f_{\rm ICM}/f_b=1$ here to be conservative.  Note that $\int \phi(M)\,dM=1$.

In evaluating this expression we use the Jenkins {\em et~al.} \cite{2001MNRAS.321..372J} mass function (fit to $N$-body simulations) and the Sheth \& 
Tormen mass-bias relation 
\cite{1999MNRAS.308..119S}.  The density-weighted gas temperature cannot be obtained from $N$-body results; we have used the analytic model of Reid \& 
Spergel \cite{2006ApJ...651..643R} for the SZ luminosity, which after application of their Eq.~(2) yields
\beq
T_e = 5.9\left(\frac {M_v}{10^{15}h^{-1}M_\odot}\right)^{2/3}\frac{M_v}M\;{\rm keV}.
\eeq
Note that here we use the virial mass $M_v$ based on the spherical collapse model for consistency with Reid \& Spergel \cite{2006ApJ...651..643R}, 
whereas the Jenkins {\em et~al.} \cite{2001MNRAS.321..372J} mass function uses $M$ defined based on a spherical overdensity of $180\bar\rho$.  We have 
used the NFW profile
and mass-concentration relation \cite{2003ApJ...593..272V} to convert $M$ to 
$M_v$.  The fraction $M_v/M$ is necessary since the total SZ luminosity predicted by Ref.~\cite{2006ApJ...651..643R} is actually the product $MT_e$.
The corresponding bounds from Section~\ref{ss:ang} are shown in Table~\ref{tab:tsz}.

We now turn to the evaluation of the 1- and 2a-halo contamination, from Eq.~(\ref{eq:contam-12a}).  The 2a-halo term can actually be calculated 
provided we know the background cosmology, mass function, bias, and mass-temperature relation:
\begin{widetext}
\beqa
\Delta C_l^{g\kappa}({\rm 2ah}) &=& \frac{r_{ps}(l)}{R_l}\int dF\sum_N \frac{NF^2}{\bar n_g}n_{2D}(N)n_{2D}(F)C_l(N;F)
= \frac{r_{ps}(l)}{R_l}\int d\chi\int dF\, F^2b_g\Pi(\chi)n_{3D}(F)b(F)P_{\rm lin}(k)
\nonumber \\
&=& \frac{r_{ps}(l)}{R_l}\int \frac{d\chi}{r^4} \int dM\,
\frac M{\bar\rho_0}\phi(M)b(M)f(\chi)P_{\rm lin}(k)
\left(2\bar\tau'T_{\rm CMB}\frac{f_{\rm ICM}}{f_b}
\frac{k_{\rm B}T_e(M)}{m_ec^2}\right)^2,
\label{eq:sz2a}
\eeqa
\end{widetext}
where the first equality involves use of the Limber approximation and the fact that the (2-halo) galaxy-halo cross spectrum is the product of biases 
times the linear power spectrum, and the second equality involves conversion of the integral from tSZ flux $F$ to halo mass $M$.  The 
denominator $\bar\rho_0$ is the mean matter density of the universe today.  The argument for the 1-halo term is similar except that here we actually 
need the number of galaxies per halo, and there is no factor of $C_l(N;F)$:
\beqa
\Delta C_l^{g\kappa}({\rm 1h}) &=& \frac{r_{ps}(l)}{R_l}\int \frac{d\chi}{r^2} \int dM\,
\frac M{\bar\rho_0}\phi(M)\frac{N(M)}{\bar n_g}
\nonumber \\ && \times 
\left(2\bar\tau'T_{\rm CMB}\frac{f_{\rm ICM}}{f_b}
\frac{k_{\rm B}T_e(M)}{m_ec^2}\right)^2.
\label{eq:sz1}
\eeqa
We may evaluate Eqs.~(\ref{eq:sz2a}) and (\ref{eq:sz1}) in a worst-case scenario by assuming $f_{\rm ICM}/f_b=1$ and $N(M)=N_{\rm max}$.  Not 
surprisingly, these equations are dominated by the low-redshift portion of the integral ($z<0.2$), especially for NVSS where there is no photo-$z$ cut 
(such cuts are impossible for radio continuum-selected objects). Unfortunately 
these equations lead to rather weak constraints, especially on the 1-halo term.  Therefore for the results in the table we multiply the integrand by 
$1-P_{\rm XBAC}$, where $P_{\rm XBAC}$ is the probability that a cluster with a given mass would pass the XBAC flux threshold.  This gives us the 1- 
and 2a-halo contributions from clusters whose X-ray flux is too low to have been included in XBAC; consequently our estimates of contamination should 
be added to those obtained from XBAC.  The probability $P_{\rm XBAC}$ was then determined using the XBAC flux limit and the lognormal
$P(L|M_{200})$ distribution from Ref.~\cite{2002ApJ...567..716R}.

\begin{table*}
\caption{\label{tab:tsz}Upper limits to the contamination to the galaxy-convergence spectrum from the tSZ effect.  The 2b- and 3-halo terms have been 
evaluated for the three cosmologies considered.  The contamination estimates are scaled to the Rayleight-Jeans limit; multiply by 0.72 to get 
contamination to the frequency-averaged lensing signal.  The local XBAC cluster constraints are computed using several assumptions about the radial 
profile, and several assumptions about the fluxes: pessimistic (``pess'', based on 95\%\ confidence upper limits) and central values (``cent'').}
\begin{tabular}{lclclclcl}
\hline\hline
Term & & Method & & $\Delta A$(LRG) & & $\Delta A$(QSO) & & $\Delta A$(NVSS) \\
\hline
2a halo & & fid. cos. & & $-0.0024$ & & $-0.0001$ & & $-0.0010$ \\
(nonlocal) & & high $\sigma_8$ & & $-0.0202$ & & $-0.0008$ & & $-0.0061$ \\
        & & high $\Omega_m$ & & $-0.0030$ & & $-0.0001$ & & $-0.0013$ \\
\hline
Local clusters & & XBAC-cent, $\beta$ (Coma), pointlike (all others) & & $-0.0007$ & & $+0.0007$ & & $-0.0039$ \\
Local clusters & & XBAC-pess, $\beta$ (Coma), pointlike (all others) & & $-0.0045$ & & $+0.0032$ & & $-0.0146$ \\
\hline
Term & & Method & & $|\Delta A|_{\rm max}$(LRG) & & $|\Delta A|_{\rm max}$(QSO) & & $|\Delta A|_{\rm max}$(NVSS) \\
\hline
1 halo & & fid. cos. & & $8.1\times 10^{-3}$ & & $7.5\times 10^{-3}$ & & $8.7\times 10^{-3}$ \\
(nonlocal) & & high $\sigma_8$ & & $4.1\times 10^{-2}$ & & $3.8\times 10^{-2}$ & & $4.4\times 10^{-2}$ \\
        & & high $\Omega_m$ & & $1.6\times 10^{-2}$ & & $1.5\times 10^{-2}$ & & $1.7\times 10^{-2}$ \\
\hline
2b halo & & fid. cos. & & $1.3\times 10^{-3}$ & & $1.4\times 10^{-3}$ & & $1.6\times 10^{-3}$ \\
        & & high $\sigma_8$ & & $1.2\times 10^{-2}$ & & $1.2\times 10^{-2}$ & & $1.4\times 10^{-2}$ \\
        & & high $\Omega_m$ & & $1.1\times 10^{-2}$ & & $1.0\times 10^{-2}$ & & $1.2\times 10^{-2}$ \\
\hline
3  halo & & fid. cos. & & $3.1\times 10^{-4}$ & & $2.9\times 10^{-5}$ & & $1.0\times 10^{-4}$ \\
        & & high $\sigma_8$ & & $3.0\times 10^{-3}$ & & $2.6\times 10^{-4}$ & & $1.0\times 10^{-3}$ \\
        & & high $\Omega_m$ & & $8.7\times 10^{-4}$ & & $4.9\times 10^{-5}$ & & $6.0\times 10^{-4}$ \\
\hline\hline
\end{tabular}
\end{table*}

\subsection{Summary of extragalactic foregrounds}
\label{ss:exfs}

The analysis of extragalactic foregrounds presented here has been quite long, and it is worth summarizing our major findings.  These have been that the 
1, 2a, and 2b-halo terms for the radio sources and tSZ effect could contribute at the level of up to several percent. For the infrared sources, the 
contamination $|\Delta A|$ (rescaled to the frequency-averaged map) is $<1$\%\ for each of the samples, even after adding 1, 2a, 2b, and 3-halo terms 
and the local contribution estimated from the IRAS maps. The radio and tSZ 3-halo terms are smaller ($\le 0.2$\%, after rescaling to the contamination 
in the cross-correlation with the frequency-averaged $TT$ lensing map); and the kSZ and Rees-Sciama effects are completely negligible ($\le 0.1$\%). 
This is good news, since the estimates we have used for the 3-halo terms and the kSZ and RS effects are crude (based on perturbation theory in the 
quasilinear regime) and we would not trust the theory to give a correction for them.

We now discuss the potentially significant foreground terms and the robustness of their contamination estimates:
\newcounter{fg}
\begin{list}{\arabic{fg}. }{\usecounter{fg}}
\item
For the radio sources, the contamination may be dominated by the 1+2a halo term (the uncertainty in this term is larger than the maximum effect from 
the 2b-halo term).  These have been constrained by constructing a product of frequency-difference maps $D$ (which contains a feature around each point 
source whose amplitude is proportional to the square of the flux) and cross-correlating this with the galaxies.  There is no detection of this 
correlation $C_l^{gD}$: the values shown in the bottom section Table~\ref{tab:rps} are consistent with zero.  The $2\sigma$ worst-case contamination 
is for the LRGs where $|\Delta A|$ could be as large as 0.15.
\item
For the tSZ effect from nearby clusters, the contamination was assessed primarily using the XBAC catalog.  The estimate for the former was normalized 
using the WMAP results \cite{2007ApJS..170..288H} and the result is that the contamination is at the $|\Delta A|\le 0.01$ level in the 
frequency-averaged maps.
\item
The tSZ effect from distant or faint clusters below the XBAC flux limit was constrained using theoretical arguments: for a given cosmology we can 
calculate the halo mass function and bias.  The worst contamination is for the high-$\sigma_8$ cosmology ($\sigma_8=0.92$) so we focus on this case.  
The 1-halo term in Table~\ref{tab:tsz} is the largest (up to 3\%), however this is an upper limit assuming that there are $N_{\rm max}$ galaxies in 
every massive cluster.  This may not be far from the case for NVSS, however even for massive clusters the LRG count is typically a factor of several 
smaller than $N_{\rm max}=36$ \cite{2007arXiv0706.0727H} and the photo-$z$ cuts should remove most LRGs from $z<0.2$ clusters.  Also most clusters will 
host much 
fewer than $N_{\rm max}=10$ quasars -- although one should remember that occasionally non-quasar extragalactic objects get counted as quasars due to 
failures in the photometric pipeline (e.g. the pipeline shredded the galaxy NGC4395 and some of the H$\,${\sc ii} regions were classified as quasars).
In contrast, the 2a-halo term is a best estimate: the contamination is $\sim 1.4$\%\ for the LRGs and less for the quasars and NVSS 
since they live at higher redshift where there are many fewer massive clusters.  While it has some uncertainty due to the $L_{\rm SZ}(M)$ relation, it 
does not dominate and so we believe the sum of all the terms (1, 2a, 2b, 3-halo, and XBAC) provides a reasonable upper bound on tSZ contamination.  
This is $|\Delta A|\le 0.057$ for the NVSS-TT correlation, and less for the LRGs and quasars.
\item
The arguments in Appendix~\ref{app:corr} show that the cross-correlations of different foregrounds are not important since none of the foregrounds are 
important individually.
\end{list}

\section{Discussion}
\label{sec:conc}

In this paper, we have measured the cross-correlation between three samples of galaxies (LRGs, quasars, and radio sources) and lensing of the CMB.  We 
find evidence for such a correlation at the $2.5\sigma$ level, with an amplitude of $1.06\pm 0.42$ times the expected signal for the WMAP cosmology.  
All sources of systematic error and foreground contamination are believed to be negligible compared to the statistical uncertainties.  Our 
measurement is consistent with the earlier nondetection by Hirata {\em et~al.} \cite{2004PhRvD..70j3501H} using LRGs and the more recent $3.4\sigma$ 
result by Smith {\em et~al.} \cite{2007PhRvD..76d3510S} using radio sources.

The case for lensing of the CMB is strengthened by having analyses by two different groups (Smith {\em et~al.} and us).  While the results both draw on 
WMAP and NVSS data, and hence are not independent, it is worth noting the very significant differences between the analysis procedures.  In the lensing 
reconstruction procedures, Smith {\em et~al.} used ${\bf C}^{-1}$ weighting of the WMAP data, whereas we force the reconstruction to use the same 
weight function $W_l$ at all frequencies and in regions with different noise levels in order to simplify the foreground analyses.  They also use 
autocorrelations of the same CMB map and subtract off the noise-induced biases, whereas we use cross-correlations which inherently have no noise bias.  
Our NVSS sample was constructed with a flux cut at 2.5$\,$mJy whereas Smith {\em et~al.} use the full catalog (except for masked regions). We also used 
cross-correlations with other galaxy samples to estimate the NVSS bias and redshift distribution (see Paper I), instead of fitting the auto-power 
spectrum; such cross-correlations are usually more robust, especially for a sample such as NVSS that shows large amounts of instrumental power.  While 
even large uncertainties in the redshift distribution do not alter the detection significance (in number of sigmas), without a solution to the redshift 
distribution problem there is no hope for CMB lensing in cross-correlation to ever become a useful cosmological probe.

We would also like to test for consistency between the two results.  Smith {\em et~al.} found a cross-correlation of $1.15\pm 0.34$ times the expected 
signal.  However, because their redshift distribution model is peaked at higher $z$, and they have a different fiducial cosmology (higher $\sigma_8$), 
their predicted signal amplitude is larger than ours; fitting the Smith {\em et~al.} cross-spectra $C_l^{g\kappa}$ to our theoretical prediction gives 
an amplitude of $A=1.32\pm 0.40$.  This compares well with our result of $A_{\rm NVSS}=1.11\pm 0.52$.  [The results are still not quite comparable 
because of the flux cut difference (2.5 versus 2.0 mJy), although we suspect this would not cause a huge difference because only 18\%\ of the NVSS 
catalog is at $<2.5\,$mJy, and the bias and redshift distribution of radio sources is usually believed to change very slowly with flux 
\cite{1998AJ....115.1693C}.  In particular to explain the entire difference in amplitude, the lensing cross-correlation signal $A$ from the faint 
$F<2.5\,$mJy sources would have to be a factor of $\sim 2$ higher than for the bright sources, which we consider unlikely.] The results were also 
arrived at independently: no revisions have been made to our central value since we became aware of the Smith {\em et~al.} result (although our error 
bars were not finalized at that time because we had not completed analyzing our simulations).  An obvious question in the comparison is why the Smith 
{\em et~al.} error bar is smaller than ours by a factor of 1.3.  Some of this is due to the 20\%\ higher number density in their Poisson-limited NVSS 
map due to their lower flux cut, which accounts for a factor of $\sqrt{1.2}$.  The remaining difference -- a factor of 1.2 in standard deviation -- 
must be attributable to all other differences in the analysis, such as their use of Q band and auto-correlations between WMAP maps, different 
handling of point sources, and (going the other direction) Smith {\em et~al.} inclusion of the point source and Galactic foreground systematics (which 
we find in this paper to be negligible).

Because of the different choices made in the analysis
the final signal amplitudes end up being
different and the Smith {\em et~al.} central value is higher.  This must 
be a statistical fluctuation if both methods are unbiased, unless the $b\,dN/dz$ is dramatically larger for the fainter sources.
These sorts of fluctuations can matter if one is trying to claim a detection, e.g. our measured amplitude would only give a 2.8$\sigma$ signal 
even with their errors versus their 3.4$\sigma$. However, in this paper 
we take the view that what is ultimately relevant is the power of the method 
to disciminate between the cosmological models and the advantage of 
having a higher detected amplitude can quickly be turned around if there
is an interesting class of models that predicts a larger lensing 
signal than the standard model, and which would therefore be more 
strongly ruled out if the measured amplitude is lower.  
It is for this reason that we put such an effort in determining 
redshift distribution weighted bias $b\,dN/dz$ for the samples, because 
without it no cosmological interpretation is possible, since 
variations in cosmology and in $b\,dN/dz$ will be degenerate. 

Another difference in the two analyses is in the treatment of extragalactic foregrounds.  We split the foregrounds into 1, 2a, 2b, and 3-halo terms for 
each foreground component, and considered each of the many resulting terms separately.  Smith {\em et~al.} did several tests, the most important being 
(i) a comparison of Q, V, and W band signals, and (ii) a point source bispectrum test which in our terminology accounts for the 1+2a halo terms.  
There are possible worries with these tests, e.g. in their implementation of (ii) the ``template'' bispectrum for the point sources has the same 
frequency dependence as the CMB.  Real point sources will have a different frequency dependence, however the template should still have some 
sensitivity to them and this is demonstrated explicitly by the Smith {\em et~al.} simulations.  The method also does not consider 2b and 3-halo terms, 
and both tests can miss kSZ or RS contaminants; however as we have argued here these are negligible.

We also included the SDSS LRGs and quasars.  These together (without NVSS) give an amplitude of $0.99\pm 0.56$ times the expected signal; this is a 
$1.8\sigma$ result and is consistent with the fiducial cosmology.
Thus in all cases the measured lensing amplitude is consistent with the WMAP cosmology, which can be counted as a success.  However it is not yet competitive with other 
cosmological probes.
We have constructed a likelihood function (see Appendix~\ref{app:lf}) and included it in the Markov chains of 
Paper I as a proof of principle, but it does not add much to the CMB+ISW constraints.
The real challenge for lensing of the CMB is to move beyond the ``first detection'' stage to providing interesting cosmological constraints.  
Several improvements must be made in order to bring lensing of the CMB to the leading edge of 
cosmology:
\newcounter{issues}
\begin{list}{\arabic{issues}. }{\usecounter{issues}}
\item
The most obvious issue is that CMB maps with higher resolution and lower noise than WMAP are required (one cannot improve on the sky 
coverage).  This will get better with the upcoming Planck satellite \footnote{URL: http://planck.esa.int/science-e/www/area/index.cfm?fareaid=17} and 
the ground-based Atacama Cosmology Telescope \footnote{URL: http://www.physics.princeton.edu/act/} and South Pole Telescope 
\footnote{URL: http://spt.uchicago.edu/}.
\item
Extragalactic foregrounds, while only a minor issue here, get worse as one uses the higher multipoles in the CMB.  
One possibility is the use of multifrequency information: at the higher frequencies probed by the future experiments, infrared point sources and tSZ 
have a distinctive spectral signature.  Another possibility is to use spatial information: a modified reconstruction procedure could be used to 
suppress the point source responsivity function $r_{ps}(l)$, which controls the 1- and 2a-halo contributions from point sources.  (It is however 
impossible to do a reconstruction that is insensitive to all possible forms of the 3-halo term.)  Yet another possibility, which is probably the best 
when the data become available, is to use polarization \cite{2002ApJ...574..566H, 2003PhRvD..68h3002H}.  The point sources, tSZ, and kSZ effects are 
either observed or theoretically expected to be 
much weaker in polarization than in temperature \cite{2000ApJ...530..133T, 2000ApJ...529...12H, 2007ApJS..170..335P}, and the RS effect is absent.  Also for 
randomly oriented point sources the 2b- and 3-halo contamination is exactly zero because two different sources can have no $QQ$ or $UU$ correlations, 
thus (aside from correlations of position angles of distinct sources) eliminating $r_{ps}(l)$ or using the multi-template bispectrum estimator of Smith 
{\em et~al.} \cite{2007PhRvD..76d3510S} would remove the point sources entirely.
\item
The third issue for lensing in cross-correlation is the theoretical uncertainty in the galaxy-convergence cross-spectrum.  In this paper we have used 
linear biasing, $P_{g\delta}(k)=b_gP_{\delta\delta}^{\rm lin}(k)$, and thrown out small-scale information ($k>0.1h\,$Mpc$^{-1}$) where linear biasing 
is not valid.  This must be improved in the future: especially for low-redshift mass tracers a large fraction of the information is in the nonlinear 
regime, and if one aims for constraints at the several percent level, simple ideas like linear biasing may not be adequate even at 
$k=0.1h\,$Mpc$^{-1}$.  There is also uncertainty in the redshift distribution of the galaxies.  One obvious method is to do lensing of the CMB in 
autocorrelation (with the four-point function \cite{2001PhRvD..64h3005H} or using the iterative approaches \cite{2003PhRvD..68h3002H}), although this 
will be very demanding on control of instrumental systematics.  If one chooses to go the cross-correlation route, then when sufficiently high 
signal-to-noise ratio is available it will be desirable to use halo modeling, and to better calibrate the photometric redshifts for the large 
scale structure tracers with spectroscopic surveys.  The latter will of course also be valuable to any cosmic shear program using the galaxies as 
lensing sources.
\end{list}

If these challenges can be met, weak lensing of the CMB will make the transition from being a simple consistency check of the cosmological model to a 
routinely used cosmological probe.

\begin{acknowledgments}

We thank Sudeep Das, Doug Finkbeiner, Jim Gunn, Beth Reid, Kendrick Smith, David Spergel, and Oliver Zahn for useful discussions.

Funding for the SDSS and SDSS-II has been provided by the Alfred P. Sloan Foundation, the Participating Institutions, the National Science Foundation, 
the U.S. Department of Energy, the National Aeronautics and Space Administration, the Japanese Monbukagakusho, the Max Planck Society, and the Higher 
Education Funding Council for England. The SDSS Web Site is http://www.sdss.org/.

The SDSS is managed by the Astrophysical Research Consortium for the Participating Institutions. The Participating Institutions are the American Museum 
of Natural History, Astrophysical Institute Potsdam, University of Basel, University of Cambridge, Case Western Reserve University, University of 
Chicago, Drexel University, Fermilab, the Institute for Advanced Study, the Japan Participation Group, Johns Hopkins University, the Joint Institute 
for Nuclear Astrophysics, the Kavli Institute for Particle Astrophysics and Cosmology, the Korean Scientist Group, the Chinese Academy of Sciences 
(LAMOST), Los Alamos National Laboratory, the Max-Planck-Institute for Astronomy (MPIA), the Max-Planck-Institute for Astrophysics (MPA), New Mexico 
State University, Ohio State University, University of Pittsburgh, University of Portsmouth, Princeton University, the United States Naval Observatory, 
and the University of Washington. 

The 2dF QSO Redshift Survey (2QZ) was compiled by the 2QZ survey team from observations made with the 2-degree Field on the Anglo-Australian Telescope.

The 2dF-SDSS LRG and QSO (2SLAQ) Survey was compiled by the 2SLAQ team from SDSS data and observations made with the 2-degree Field on the
Anglo-Australian Telescope.

This publication makes use of data products from the Two Micron All Sky Survey, which is a joint project of the University of Massachusetts and the 
Infrared Processing and Analysis Center/California Institute of Technology, funded by the National Aeronautics and Space Administration and the 
National Science Foundation.

We acknowledge the use of the Legacy Archive for Microwave Background Data Analysis (LAMBDA). Support for LAMBDA is provided by the NASA Office of 
Space Science.  Some of the results in this paper have been derived using the HEALPix \cite{2005ApJ...622..759G} package.

C.H. was a John Bahcall fellow at the Institute for Advanced Study during most of the preparation of this paper.
N.P. is supported by a Hubble Fellowship
HST.HF-01200.01 awarded by the Space Telescope Science Institute,
which is operated by the Association of Universities for Research in
Astronomy, Inc., for NASA, under contract NAS 5-26555. Part of this work was supported by the
Director, Office of Science, of the U.S.
Department of Energy under Contract No. DE-AC02-05CH11231.
U.S. is supported by the Packard Foundation and NSF
CAREER-0132953.

\end{acknowledgments}

\appendix

\section{Contamination from extragalactic microwave sources}
\label{app:cont}

The purpose of this appendix is to examine how extragalactic sources (radio or infrared point sources, or Sunyaev-Zel'dovich effect) affect the 
measurement of the galaxy-convergence cross-power $C_l^{g\kappa}$.

From Eqs.~(\ref{eq:glm}) and (\ref{eq:v}), we can see that after averaging over the WMAP instrument noise we have contamination
\beqa
\langle\Delta v_{lm}^{(\parallel)}\rangle &=& (-1)^m (2l+1)
\sum_{l'l''m'm''} {\cal F}_{ll'l''}
\nonumber \\ && \times
\threej{l}{l'}{l''}{-m}{m'}{m''}T_{l'm'}T_{l''m''},
\label{eq:vv}
\eeqa
where we have introduced the coupling coefficient
\beq
{\cal F}_{ll'l''} = \frac{\sqrt{l(l+1)}}{4(2l+1)C_{l'}^{\rm wt}C_{l''}^{\rm wt}}{\cal J}_{ll'l''}
e^{-l(l+1)\sigma_0^2/2}.
\label{eq:f}
\eeq

The three-mode correlation function of the galaxies and the temperature is defined via the relation
\beq
\langle g_{lm}T_{l'm'}T_{l''m''} \rangle = B_{ll'l''}^{gTT} \threej{l}{l'}{l''}{m}{m'}{m''}.
\label{eq:xtt}
\eeq
Combining with Eq.~(\ref{eq:vv}), and using the reality condition $v_{lm}^{(\parallel)\ast}=(-1)^mv_{lm}^{(\parallel)}$, we find
\beqa
\Delta C_l^{g\kappa} &=& \frac 1{R_l}\langle g_{lm}\Delta v_{lm}^{(\parallel)\ast} \rangle
\nonumber \\
&=& \frac{2l+1}{R_l} \sum_{l'l''m'm''}
 {\cal F}_{ll'l''}
\nonumber \\ && \times
\threej{l}{l'}{l''}{m}{m'}{m''}\langle g_{lm}T_{l'm'}T_{l''m''}\rangle
\nonumber \\
&=& \frac{2l+1}{R_l} \sum_{l'l''m'm''} {\cal F}_{ll'l''}
\threej{l}{l'}{l''}{m}{m'}{m''}^2 B_{ll'l''}^{gTT}
\nonumber \\
&=& \frac 1{R_l}\sum_{l'l''} {\cal F}_{ll'l''} B_{ll'l''}^{gTT},
\label{eq:dcl}
\eeqa
where in the last line we have used the normalization of the $3j$ symbol to collapse the sum over azimuthal quantum numbers.  Therefore any bias 
produced by point sources in the bispectrum $B_{ll'l''}^{gTT}$ produces a corresponding bias in the galaxy-convergence cross-correlation.  This is 
expressed in Eq.~(\ref{eq:dcl-text}).

\section{Derivation of terms in point source bispectrum}
\label{app:bs}

This appendix derives the various multi-halo terms in the galaxy-point source-point source bispectrum, used in Section~\ref{ss:ang}.  In each case, the 
starting point for the derivation is Eq.~(\ref{eq:xtt}).

\subsection{1-halo term}

The 1-halo term is the simplest to derive.  If a halo containing $N$ galaxies with flux $F$ is found at position $\nhat$, then the contribution to 
$g_{lm}$ is $(N/\bar n_g)Y_{lm}^\ast(\nhat)$ and the contribution to $T_{lm}$ is $FY_{lm}^\ast(\nhat)$.  The 1-halo contribution to the 3-mode 
correlation function is then obtained by integrating (or summing) over $F$, $N$, and $\nhat$:
\beqa
\langle g_{lm}T_{l'm'}T_{l''m''} \rangle_{\rm 1h} &=& \int dF\sum_N\int d^2\nhat
\frac{N}{\bar n_g}F^2
\nonumber \\ && \times
Y_{lm}^\ast(\nhat) Y_{l'm'}^\ast(\nhat) Y_{l''m''}^\ast(\nhat)
\nonumber \\
&=& \int dF\sum_N \frac{N}{\bar n_g}F^2
\eta_{ll'l''}
\nonumber \\ && \times
\threej{l}{l'}{l''}{m}{m'}{m''},
\label{eq:ap-1}
\eeqa
where $\eta_{ll'l''}$ is the combination of coefficients in Eq.~(\ref{eq:etadef}).  In the second equality we have used the three spherical harmonic 
integral formula, and removed the conjugation symbol because the $3j$ symbol is real.  Comparison to Eq.~(\ref{eq:xtt}) proves Eq.~(\ref{eq:b1}).

\subsection{2a-halo term}

The 2a-halo term involves two haloes, one with $N$ galaxies, and one with flux $F$.  These two types of halo have an angular power 
spectrum $C_l(N;F)$ and correspondingly have an angular correlation function $w(\theta|N;F)$.
The probability density to find these two haloes at locations $\nhat$ (for the halo containing the galaxy) and $\nhat'$ (for the halo containing the 
point source) is then modulated by the factor $1+w(\theta)$:
\beqa
\!\!\!\!\!\!\!\!\!\langle g_{lm}T_{l'm'}T_{l''m''} \rangle_{\rm 2ah}\!\!&=& \int dF\sum_N\int d^2\nhat\int d^2\nhat'
\nonumber \\ && \times
\frac{N}{\bar n_g}F^2[1+w(\theta)]
Y_{lm}^\ast(\nhat) 
\nonumber \\ && \times
Y_{l'm'}^\ast(\nhat') Y_{l''m''}^\ast(\nhat')
\label{eq:a-2a-1}
\eeqa
where $\theta$ is the angle between $\nhat$ and $\nhat'$.  This factor can be re-written as
\beq
1+w(\theta) = \sum_{LM} [4\pi\delta_{L0}+C_L(N;F)]Y_{LM}(\nhat) Y_{LM}^\ast(\nhat').
\eeq
(The usual expression in terms of Legendre polynomials of $\cos\theta$ is equivalent to 
this by the spherical harmonic addition theorem.)  Then we have
\beqa
\!\!\!\!\langle g_{lm}T_{l'm'}T_{l''m''} \rangle_{\rm 2ah}
\!\!&=&
\sum_{LM}\int dF\sum_N \frac{N}{\bar n_g}F^2
\nonumber \\ && \times
[4\pi\delta_{L0}+C_L(N;F)]
\nonumber \\ && \times
\int d^2\nhat\, Y_{LM}(\nhat)Y_{lm}^\ast(\nhat)
\nonumber \\ && \times
\int d^2\nhat'\, Y_{LM}^\ast(\nhat')
\nonumber \\ && \times
Y_{l'm'}^\ast(\nhat') Y_{l''m''}^\ast(\nhat').
\eeqa
The $\nhat$ integral is collapsed using spherical harmonic orthonormality to $\delta_{Ll}\delta_{Mm}$, and the $\nhat'$ integral is a 3-harmonic 
integral.  Therefore for $l\neq 0$ we have
\beqa
\langle g_{lm}T_{l'm'}T_{l''m''} \rangle_{\rm 2ah}
\!\!&=&
\int dF\sum_N \frac{N}{\bar n_g}F^2 C_l(N;F)
\nonumber \\ && \times
\eta_{ll'l''} \threej{l}{l'}{l''}{m}{m'}{m''}.
\eeqa
The evenness of $l+l'+l''$ for nonzero $\eta_{ll'l''}$ and reality of the $3j$ symbol are used here, just as in Eq.~(\ref{eq:ap-1}).  This equation 
proves Eq.~(\ref{eq:b2a}).

\subsection{2b-halo term}

Here we consider the $gTT$ bispectrum introduced by correlations of a halo at $\nhat$ with $N$ galaxies emitting flux $F$, and a halo at $\nhat'$ 
emitting flux $F'$.  The angular power spectrum of these two haloes is $C_l(N,F;F')$.  For this case, we have in analogy to Eq.~(\ref{eq:a-2a-1}):
\beqa
\!\!\!\!\!\!\!\!\!\langle g_{lm}T_{l'm'}T_{l''m''} \rangle_{\rm 2bh}\!\!&=& \int dF\int dF'\sum_N\int d^2\nhat\int d^2\nhat'
\nonumber \\ && \times
\frac{N}{\bar n_g}FF'[1+w(\theta)]
Y_{lm}^\ast(\nhat)
\nonumber \\ && \times
Y_{l'm'}^\ast(\nhat) Y_{l''m''}^\ast(\nhat')
\nonumber \\ && + (l'm'\leftrightarrow l''m'').
\eeqa
The symmetrization in the last lie occurs because the halo at $\nhat$ could contribute to $T_{l'm'}$ and that at $\nhat'$ could contribute to 
$T_{l''m''}$, or vice versa.  Manipulations similar to those for the 2a-halo term then lead to Eq.~(\ref{eq:b2b}).

\section{Small extragalactic foregrounds}
\label{app:efg}

\subsection{Kinetic SZ effect}
\label{ss:ksz}

The galaxy-kSZ-kSZ bispectrum is another potential contaminant of the CMB lensing signal, and is especially worrying for future experiments because the 
kSZ signal has the same frequency dependence as CMB lensing \cite{2004NewA....9..687A}.  This section constructs a rough upper limit to the kSZ 
contamination of the lensing signal, and finds that this contamination is negligible in WMAP.

The kSZ temperature anisotropy is a line-of-sight integral,
\beq
T_{\rm kSZ} = \frac{T_{\rm CMB}}c \int d\chi\,\bar\tau' (1+\delta_b) u_{b,\parallel},
\eeq
where $\delta_b$ and $u_b$ are the baryon density perturbations and velocity $\parallel$ denotes the line-of-sight component of velocity (positive 
toward the observer), and 
$\bar\tau'$ is the mean Thomson optical depth per comoving radial 
distance.  The velocity field will be irrotational (except for contributions on very small scales due to nonadiabatic effects) so the integral of 
velocity cancels along the line of sight.  We will also make the approximation that $\delta_b$ is equal to the matter density $\delta$, which should 
again be valid except on very small scales.  (For comparison, the dominant contribution to the lensing signal comes from $l\sim 400$, or 
$k=0.4h\,$Mpc$^{-1}$ at the 25th percentile redshift of the LRGs.)  In this case, the kSZ temperature increment becomes $c^{-1} \int d\chi\,\bar\tau' 
\delta\,u_\parallel$ and the galaxy-kSZ-kSZ bispectrum is given by
\beqa
B_{ll'l''}^{gTT}({\rm kSZ}) &=& \frac{\eta_{ll'l''}T_{\rm CMB}^2}{c^2}
\int \frac{d\chi}{r^4}{\bar\tau'}{^2}\Pi(\chi)
\nonumber \\ && \times
B(k,k',k''|g;\delta\,u_{\parallel};\delta\,u_{\parallel}).
\label{eq:l-ksz}
\eeqa
Here $\Pi(\chi)$ is the comoving distance distribution of the galaxies.

In order to proceed we make a further assumption that the velocity field is perfectly coherent, i.e. that all of the power in $u_{\parallel}$ is at 
very small wavenumber and we can obtain the statistics of the kSZ signal by multiplying the statistics of the density field by the appropriate power of 
some large-scale velocity $u_{\parallel}$ and then averaging (this gives $u_{\rm rms}^2/3$ for statistics involving two copies of the kSZ signal; the 
factor of 1/3 comes from the fact that the kSZ effect depends on only 1 of the 3 components of velocity).  This 
idealization is commonly made in predictions of the small-scale kSZ power spectrum and should be valid for predictions on scales smaller than the peak 
of the velocity power, i.e the maximum of $kP(k)$ at $k\sim 0.05h\,$Mpc$^{-1}$.  Since larger scales are considered here, especially for the 
high-redshift part of the integral relevant to the quasars and NVSS, this assumption is probably {\em not} valid -- instead the velocity field will 
become incoherent and the radial velocities of two widely separated regions will have only a small correlation.  This is not a problem for us since the 
loss of coherence suppresses the kSZ signal, i.e. we maximize the possible kSZ signal by ignoring it.  The reader should thus keep in mind that the 
galaxy-kSZ-kSZ bispectrum derived here may be a large overestimate.  We have thus converted Eq.~(\ref{eq:l-ksz}) into
\beqa
B_{ll'l''}^{gTT}({\rm kSZ}) &=& \eta_{ll'l''}T_{\rm CMB}^2 \frac{u_{\rm rms}^2}{3c^2}
\int \frac{d\chi}{r^4} {\bar\tau'}{^2}\Pi(\chi)
\nonumber \\ && \times
B(k,k',k''|g;\delta;\delta).
\eeqa
The simplest way to calculate the bispectrum is to use second-order perturbation theory assuming linear bias for the galaxies, i.e.
\beqa
B_{ll'l''}^{gTT}({\rm kSZ}) &\approx& \eta_{ll'l''} \frac{u_{\rm rms}^2}{3c^2}T_{\rm CMB}^2
\int \frac{d\chi}{r^4} {\bar\tau'}{^2}
\nonumber \\ && \times
f(\chi) B_{\rm 2P}(k,k',k''),
\eeqa
where $f(\chi)$ is the product of the bias and comoving distance distribution.  The calculation of the implied contamination to the lensing signal can 
be carried out just as for the point source 3-halo term (Eq.~\ref{eq:bound3}): one only makes the replacement
\beq
b_T\bar T' \rightarrow \frac{u_{\rm rms}}{\sqrt3\,c} \bar\tau'\,T_{\rm CMB}.
\eeq
Just as in Eq.~(\ref{eq:bound3}), we take the absolute value of each term in the sum over $l'$ and $l''$ to avoid accidental cancellation of terms due 
to the configuration dependence of the bispectrum (which for this extremely crude calculation we do not trust), and use linear theory to obtain 
$u_{\rm rms}$.  For the fiducial cosmology, linear theory gives $u_{\rm rms}=430\,$km$\,$s$^{-1}$ today.  The resulting maximum contamination to the 
lensing signal is found to be $|\Delta A|\le 10^{-3}$ for each of the three cosmologies and each sample.  The calculation presented here is very crude, 
nevertheless it suffices to establish that the kinetic Sunyaev-Zel'dovich effect has no significant impact on our results.

\subsection{ISW/Rees-Sciama effect}
\label{ss:rs}

Another foreground signal is the Rees-Sciama (RS) effect, i.e. the sourcing of temperature fluctuations by changing gravitational potentials.  This is 
physically the same phenomenon as the ISW effect, although the label ``ISW'' is usually used to refer to the linear-regime effect; for brevity we will 
use the label RS here to refer to the effect on all scales.  It arises at low redshift and thus can produce a galaxy-RS-RS bispectrum.  This section 
considers an order-of-magnitude estimate of the galaxy-RS-RS bispectrum and the consequent contamination to the lensing signal, and shows that it is 
negligible.

The RS effect can be written as
\beq
T_{\rm RS} = 2\frac{T_{\rm CMB}}{c^3} \int d\chi\,\dot\Psi,
\eeq
where $\Psi$ is the dimensionless gravitational potential and the dot denotes derivative with respect to conformal time.  It is related to the density 
field via the Poisson equation,
\beq
\Psi({\bf k}) = \frac{3\Omega_mH_0^2}{2ac^2k^2} \delta({\bf k}).
\label{eq:deltapsi}
\eeq
For linearly biased galaxies, the galaxy-RS-RS bispectrum can then be written as
\beq
B^{gTT}_{ll'l''}({\rm RS}) = 4\frac{T_{\rm CMB}^2}{c^2} \eta_{ll'l''}\int \frac{d\chi}{r^4} f(\chi) B(k,k',k''|\delta;\dot\Psi;\dot\Psi).
\label{eq:b-rs}
\eeq
The problem is then to estimate the 3-dimensional bispectrum $B(k,k',k''|\delta;\dot\Psi;\dot\Psi)$.
All bispectra are zero in linear perturbation theory with Gaussian initial conditions, so we have used the tree-level second-order bispectrum.
The contamination to the galaxy-convergence cross-spectrum can then be estimated using a method analogous to Eq.~(\ref{eq:dcl-text}):
\beq
|\Delta C_l^{g\kappa}| \le R_l^{-1}\sum_{l'l''} | {\cal F}_{ll'l''}B^{gTT}_{ll'l''}({\rm RS}) |.
\eeq
As before, the absolute values are used to prevent accidental cancellation of terms due to the configuration dependence of $B^{gTT}_{ll'l''}$(RS) -- 
once again we do not trust the above calculation to give the configuration dependence correctly.  The worst contamination found for any of the 3 
samples and 3 models is $|\Delta A|\le1.1\times 10^{-3}$.
While the calculation presented is obviously very rough, these numbers do establish that the ISW/Rees-Sciama effect 
is not a significant contaminant.

\section{Correlations between different foregrounds}
\label{app:corr}

In the ISW analysis, for which we measure $C_l^{gT}$, the foregrounds added linearly in the sense that $C_l^{gT}$ is the sum of the ``signal'' (ISW) 
plus the contribution from point sources, plus the contribution from Galactic emission, etc.  Because lensing depends on the bispectrum 
$B_{ll'l''}^{gTT}$, the foregrounds no longer add linearly: in addition to lensing, LSS-point source-point source, and LSS-Galactic-Galactic terms, it 
is possible to have cross-terms involving multiple foregrounds.  Out of the 6 foregrounds we have considered -- Galactic emission, radio sources, 
infrared sources, tSZ, kSZ, and RS -- it is possible to construct $6\times 5/2=15$ cross-bispectra of the type LSS-foreground~\#1-foreground~\#2.  
Not all 10 combinations are possible: physically the foreground structure of the Galaxy cannot correlate with extragalactic foregrounds, and kSZ 
exhibits a reversal of sign depending on the radial velocity that is not exhibited by point sources, tSZ, or RS, so correlations involving these 
two foregrounds are not possible.  This leaves us with 6 possible foreground correlations involving radio point sources (rps), infrared point sources 
(irps), tSZ, and RS.  As we have seen the RS effect is small and we will not consider its cross-correlations with other foregrounds here.

\subsection{Galaxy-radio source-infrared source}

The correlated-foreground signals involving point sources are most easily investigated in the context of the halo model.  We investigate it first at a 
single frequency and then consider what happens in a multifrequency analysis such as ours.

The 1- and 2a-halo terms for the total foreground contain a factor of $F^2$ in Eq.~(\ref{eq:contam-12a}) where $F$ is the flux from that source.  In 
the presence of multiple 
emission components (rps, irps, tSZ) these terms contain the square of the total flux, i.e.
\beq
F_{\rm tot}^2 = \sum_i F_i^2 + \sum_{i<j} 2F_iF_j.
\eeq
Therefore the contamination from galaxy-rps-irps is obtained from Eq.~(\ref{eq:contam-12a}) by the replacement $F^2\rightarrow 2F_{\rm rps}F_{\rm 
irps}$.  Now the 1+2a-halo corrections are all semipositive-definite linear combinations of $F^2$'s, i.e. we may schematically write
\beq
-\Delta A = \sum_\mu k_\mu F^{(\mu)}_{\rm tot}{^2}
\eeq
where $k_\mu\ge 0$ and the sum is over halo types $\mu$.  (The $-$ sign is introduced since the contributions to $\Delta A$ are negative, as we are 
using only multipoles with the negative sign of $r_{ps}(l)/R_l$ and assuming
positively correlated haloes.  Also in the context of the usual halo model the sum would actually be an integral over the halo mass, as well as any 
additional relevant parameters; this does not affect the validity of the following argument.)  However a trivial application of Cauchy's inequality 
(see Eq.~3.2.9 of Ref.~\cite{1972hmf..book.....A}) implies that
\beq
\sum_\mu k_\mu 2F^{(\mu)}_{\rm rps} F^{(\mu)}_{\rm irps} \le 2\sqrt{\left[\sum_\mu k_\mu F^{(\mu)}_{\rm rps}{^2}\right]
\left[\sum_\mu k_\mu F^{(\mu)}_{\rm irps}{^2}\right]}.
\label{eq:2f}
\eeq
The qualitative conclusion is that the cross-talk between radio and infrared sources must be negligible if both types of sources are 
individually negligible; the quantitative result is that -- since we only use multipoles with the same sign of $r_{ps}(l)/R_l$ -- 
Eqs.~(\ref{eq:contam-12a}) and (\ref{eq:2f}) imply
\beq
|\Delta A|_{{\rm rps}\times{\rm irps}}({\rm 1,2a}) \le 2\sqrt{|\Delta A|_{\rm rps}({\rm 1,2a})
|\Delta A|_{\rm irps}({\rm 1,2a})}.
\label{eq:cross-contam}
\eeq
(We have written the 1+2a halo term with 1,2a in parentheses instead of 1+2ah to save space.)

The derivation of Eq.~(\ref{eq:cross-contam}) is valid only at one frequency, i.e. for contamination to the galaxy-VV or galaxy-WW correlations.  If 
one makes no assumptions about the SED of the sources, then it does {\em not} need to be valid for the galaxy-VW correlation.  [A simple counterexample 
is that if one decided to make the V-band flux of all infrared sources go to zero while keeping their W-band luminosity fixed, then the left-hand side 
of Eq.~(\ref{eq:cross-contam}) stays finite because of the galaxy-radio source (V band)-infrared source (W band) term, but the right side becomes zero 
since there is no infrared source contamination to the galaxy-$T_V$-$T_W$ bispectrum.]  However because our weights for the ``TT'' frequency 
combination $a_{VV}:a_{VW}:a_{WW}$ are approximately in the ratio $0.53^2:2(0.53)(0.47):0.47^2$, we should be able to approximate the galaxy-TT 
contamination by the replacing $F$ in the above equations with $0.53F_V+0.47F_W$.  The constraints such as Eq.~(\ref{eq:cross-contam}) are then valid 
for the TT case.

\subsection{Galaxy-point source-tSZ}

Essentially the same arguments of the previous section can be made to the galaxy-point source-tSZ correlation.  The one exception is that the tSZ flux 
is negative whereas the point source flux is positive.  Thus for the 1+2a-halo term Eq.~(\ref{eq:cross-contam}) becomes
\beqa
|\Delta A|_{{\rm ps}\times{\rm tSZ}}^{VV}({\rm 1,2a}) &\le& 2\sqrt{|\Delta A|^{VV}_{\rm ps}({\rm 1,2a})
|\Delta A|^{VV}_{\rm tSZ}({\rm 1,2a})}
\nonumber \\
&\le& |\Delta A|^{VV}_{\rm ps}({\rm 1,2a}) + |\Delta A|^{VV}_{\rm tSZ}({\rm 1,2a}).
\nonumber \\ &&
\label{eq:cross-contam-sz}
\eeqa
In this case, because $\Delta A_{{\rm ps}\times{\rm tSZ}}$ has the opposite sign as $\Delta A_{\rm ps}$ or $\Delta_{\rm tSZ}$ (it is positive instead 
of negative) it follows from this that the galaxy-ps-tSZ contamination has the opposite sign as the galaxy-ps-ps and galaxy-tSZ-tSZ contamination, but 
it must be smaller in magnitude than their sum.  In other words, inclusion of the 1+2a-halo galaxy-ps-tSZ term can only reduce the foreground 
contamination, and we can be conservative by ignoring it.

Once again, a similar argument applies to the 2b- and 3-halo terms, and the result should be valid for TT as well as for VV or WW.

\section{Likelihood function}
\label{app:lf}

The weak lensing of the CMB does not yet provide a competitive cosmological constraint.  Nevertheless we have included it in some of
the Markov chains in Paper I, both for completeness and as a proof of principle.  This section describes the lensing likelihood function.

We have included only the amplitude information from the galaxy-convergence cross-spectrum $C_l^{g\kappa}$ (i.e. not the shape) since the
signal-to-noise is not high enough to reliably determine the latter.  The amplitude is given by three numbers $A$(LRG), $A$(QSO), and $A$(NVSS)
measured in this paper, each of which is the amplitude for one of the samples normalized to $A=1$ for the fiducial WMAP cosmology.  In order to 
construct
a likelihood function, we need (i) the best values of each $A^\mu$ where $\mu\in$\{LRG,QSO,NVSS\}; (ii) their covariance matrix $C^{\mu\nu}$; and (iii)
a function to compute the expected value of $A^\mu$ for any choice of cosmological parameters ${\bf p}$.  Then we have a $\chi^2$ function
\beq
\chi^2 = [{\bf C}^{-1}]_{\mu\nu} [A^\mu({\rm obs})-\langle A^\mu\rangle_{({\bf p})}]
[A^\nu({\rm obs})-\langle A^\nu\rangle_{({\bf p})}].
\eeq
Items (i) and (ii) are easy to come by: the observed values of $A$ are $0.727$, $1.200$, and $1.110$, and the inverse-covariance matrix is
\beq
{\bf C}^{-1} = \left\{\begin{array}{rrr}
 1.951 & -0.263 & -0.900 \\
-0.263 &  1.871 &  0.110 \\
-0.900 &  0.110 &  4.025
\end{array}\right\}.
\eeq
Item (iii) is slightly harder.  For each cosmology and each sample we can calculate the cross-power spectrum $C_l^{g\kappa}$ just as was done for the
fiducial model in Eq.~(\ref{eq:clxk}).  After this we must determine the expectation values of the cross-spectrum estimators $\hat
c^A\equiv\hat C_{l_A}^{g\kappa}$ constructed in Section~\ref{ss:cs}.  We can construct this by recalling that $\hat c^A=[{\bf F}^{-1}]^{AB}q_B$
where ${\bf F}$ is the response matrix and ${\bf q}$ is the vector of quadratic estimators. Then
\beqa
\langle g_iv_j\rangle &=& \sum_l R_lC_l^{g\kappa}\Pi_{lij},
\nonumber \\
\Pi_{lij} &=& \sum_{m=-l}^l Y_{lm}^\ast(i)Y_{lm}^\parallel(j),
\eeqa
where ${\bf g}$ is the length-$N_{\rm pix,LSS}$ vector of galaxy overdensities in each pixel and ${\bf v}$ is the length-$2N_{\rm pix,CMB}$ vector
corresponding to the reconstructed lensing map (${\bf v}$ is a vector field on the sphere so there are 2 components per pixel).  Then we have
\beqa
\langle q_A\rangle &=& \langle {\bf g}^T{\bf w}^{(g)}{\bf P}_A{\bf w}^{({\bf v})}{\bf v}\rangle
\nonumber \\
&=& \sum_l R_lC_l^{g\kappa} {\rm Tr}\,\left[ {\bf w}^{(g)}{\bf P}_A{\bf w}^{({\bf v})}\Pi_l^T \right].
\eeqa
It follows that the expectation values of the cross-spectrum estimators are
\beq
\langle \hat C_{l_A}^{g\kappa}\rangle = \sum_{l'} W_{l'}^A C_{l'}^{g\kappa},
\eeq
where the quadratic estimator window function is
\beq
W^A_{l'} = R_{l'} [{\bf F}^{-1}]^{AB}{\rm Tr}\,\left[ {\bf w}^{(g)}{\bf P}_B{\bf w}^{({\bf v})}\Pi_{l'}^T \right].
\eeq
From Eq.~(\ref{eq:a}) we then see that
\beq
\langle A\rangle_{({\bf p})} = \sum_{l'}
 \frac{\sum_{AB}[{\bf C}_w]^{-1}_{AB}C_{l_A}^{g\kappa}({\rm fid})W_{l'}}
{\sum_{AB}[{\bf C}_w]^{-1}_{AB}C_{l_A}^{g\kappa}({\rm fid})C_{l_B}^{g\kappa}({\rm fid})}
C_{l'}^{g\kappa},
\eeq
where $C_{l_A}^{g\kappa}({\rm fid})$ is the theoretical cross-spectrum for the fiducial cosmology.  In this equation, only $C_{l'}^{g\kappa}$ needs to
be recomputed for each new cosmology; the remaining coefficients can be computed once and saved.

\bibliography{lens3}

\end{document}